\documentclass[traditabstract,twocolumn]{aa}
\usepackage{graphicx}
\usepackage{txfonts}
\usepackage{aalongtable, lscape}
\usepackage{natbib}
\begin{document}
\title{The globular cluster system of NGC1316. III. Kinematic complexity    \thanks{Based
    on observations obtained with the VLT  at ESO,  Cerro Paranal, Chile under the programme 078.B-0856}}

\subtitle{}
\titlerunning{Globular cluster system of  NGC 1316}

\author{T. Richtler
              \inst{1} 
	  \and M. Hilker
           \inst{2} 
	  \and B. Kumar 
	      \inst{3} 
	  \and L.P. Bassino 
           \inst{4}
           \and M. G\'omez
           \inst{5} 
         \and B. Dirsch
            \inst{6}  
           }
\offprints{T. Richtler}

\institute{
             Departamento de Astronom\'{\i}a, Universidad de Concepci\'on,  Casilla  160-C, Concepci\'on, Chile
              \and 
	European Southern Observatory, Karl-Schwarzschild-Str.~2,
                D-85748 Garching, Germany 
               \and
                Aryabhatta Research Institute of Observational Sciences,
                Manora Peak, 263129 Nainital, India 
                \and
                Grupo de Investigaci\'on CGGE, Facultad de Ciencias Astron\'omicas y
Geof\'isicas, Universidad Nacional de La Plata, and 
Instituto de Astrof\'isica de La Plata (CCT La Plata -- CONICET, UNLP),
Paseo del Bosque S/N, B1900FWA La Plata, Argentina
               \and
               Departamento de Ciencias  F\'{\i}sicas, Facultad de Ciencias Exactas,
               Universidad Andres Bello, Santiago, Chile
                          \and
         Friedrich-Ebert Gymnasium, Ollenhauerstr.5,
         53113 Bonn, Germany
}
%Departamento de Astronom\'{\i}a,
%Universidad de Concepci\'on,
%Concepci\'on, Chile;
%[tom]@astro-udec.cl
%\and
%Aryabhatta Institute of Observational Sciences (ARIES), Nainital, India

\date{Received  / Accepted }

\abstract{
The merger remnant NGC 1316 (Fornax A) is one of the most important objects 
regarding the investigation of and thus an important object to study 
merger-related processes.
A recent photometric study  used globular clusters in NGC 1316 to constrain 
its star formation history, but without the knowledge of individual radial 
velocities. 
The kinematical properties of the globular cluster system in comparison with the diffuse stellar light
might reveal more insight into the formation of NGC 1316. 
Of particular interest is the dark matter content. Planetary nebulae in NGC 1316 indicate a massive dark halo, and globular cluster  velocities provide independent evidence. 
We aim at measuring radial velocities of globular clusters in NGC 1316. We use these kinematical data to investigate the global structure of NGC 1316 and to constrain the
dark matter content.
We perform multi-object-spectroscopy with VLT/FORS2 and MXU. Out of 562 slits,  we extract radial velocities for 177 globular clusters.  Moreover, we measure
radial velocities of the integrated galaxy light, using slits with a sufficiently bright "sky". To these data, we add 20  cluster  velocities from Goudfrooij et al. 2001b. In an appendix,
we identify new morphological features of NGC 1316 and its companion galaxy NGC 1317.
The GC sample based on radial velocities confirms the colour peaks already found in our photometric study. The bright clusters, which  probably have their origin in a 2 Gyr-old starburst and younger star formation events,
  avoid the systemic velocity.   A Gaussian velocity distribution is found only for  clusters fainter than about $m_R=22$ mag. The velocity distribution
  of clusters shows  a pronounced peak at 1600 km/s. These clusters populate a wide area in the south-western region which we suspect to be a disk population.
  Globular clusters or subsamples of them do not show a clear rotation signal. This is  different from the galaxy light, where rotation along the major axis is
  discernable out to 3\arcmin\ radius. 
  The kinematic major axis of NGC 1316 is misaligned by about 10$^\circ$ with the photometric major axis, which might indicate a triaxial symmetry.  A simple spherical model like that suggested by  dynamical analyses of planetary nebulae
 reproduces also the velocity dispersions of the faint globular clusters. 
The  central dark matter density of  the present model  resembles a giant elliptical galaxy. This contradicts population properties which
indicate spiral galaxies as pre-merger components.  MOND would provide a solution, but the kinematical
complexity of NGC 1316 does  not allow a really firm conclusion. However, NGC 1316
might anyway be a problem for a CDM scenario, if the high dark matter density in the inner region is
confirmed in future studies. 
}


\maketitle

\section{Introduction}

Globular clusters are "Guides to galaxies" \citep{richtler09}. The photometric and kinematic properties of a globular cluster system (GCS) permit to identify subpopulations, to constrain scenarios of galaxy formation and star formation history, and to discover the existence and shape of
dark matter halos at large galactocentric radii (see \citealt{brodie06} and \citealt{harris10} for  reviews).  Most kinematical studies of GCSs targeted old
elliptical galaxies (e.g. \citealt{strader11,schuberth10,lee10,romanowsky09,lee08,schuberth06,richtler04} and references therein) whose GCSs  show the typical "bimodality" of blue and red clusters   which also corresponds
to different kinematical properties (e.g. \citealt{schuberth10}).   The red ({\it bona fide} metal-rich) clusters might have been formed in the star-burst which
formed the main body  of elliptical galaxies, while the blue ({\it bona fide} metal-poor) population might have been donated mainly by the infall
of dwarf galaxies (see \citealt{richtler13} for an overview). The rich GCSs of giant ellipticals, which are dominated by the metal-poor cluster populations, have  probably been shaped
by secular evolution rather than by major merger events \citep{vandokkum10,genel08}.

%The  GCSs of merger remnants are poorly investigated. 
The target of the present contribution is the GCS of NGC 1316 (Fornax A). This galaxy is a well investigated merger remnant in the outskirts of the Fornax cluster and
has been studied in many wavelength domains from the X-ray to the radio.  See \citet{richtler12a}(hereafter Paper I) for a
representative summary of the work done on NGC 1316.  To that we add the work on planetary nebulae by \citet{mcneil12} and on the mass functions of star clusters by \citet{goudfrooij12}. 

 In the Washington photometry from Paper I, the GCS of NGC 1316 appears quite different from that of
a giant elliptical galaxy. The blue and red GCs of giant ellipticals show peaks with Washington colours at C-R=1.35 and C-R=1.75, respectively 
\citep{bassino06b}.

 The colour distribution  of bright clusters in the bulge region shows a clear bimodality,
which, however, has a meaning different from that in giant ellipticals. A peak at C-R=1.4 marks a starburst with an age of about 2 Gyr, which
is in agreement with the spectroscopic ages and abundances  of three massive star clusters \citep{goudfrooij01b}. A bluer peak at C-R=1.1 probably
corresponds to a more recent starburst 0.8 Gyr ago, but spectroscopic confirmation is still pending. 
 A small sample of 22 bright GCs in the innermost region with radial velocities already exists \citep{goudfrooij01b}. In this paper
we present the radial velocities of 172 additional objects in a wide field. We describe the kinematics of this cluster sample and also make dynamical
remarks on this complex system.

This paper is the third in a series devoted to the cluster system of NGC 1316. 
%Paper I  presents a wide-field photometric
%study in the Washington system, while
 Paper II \citep{richtler12b} investigates the remarkable object SH2, perhaps a dwarf galaxy which recently formed a cluster population.
%Paper IV will present the dataset and remarks on the observations and reduction.

 We adopt a distance of 17.8 Mpc, quoted by \citet{stritzinger10} using the four type Ia supernovae which appeared so far in NGC 1316. At this distance, 1\arcsec
 corresponds to 86.3 pc. The heliocentric systemic velocity of NGC 1316 is  1760 km/s \citep{longhetti98}

 \section{Observations and Data Reduction}
\subsection{Observations}

The observations  were performed in service mode during  seven nights (period November 14th to December 21th
2006) at the European Southern Observatory (ESO) Very Large Telescope
(VLT) facility at Cerro Paranal, Chile (programme 078.B-0856(A), PI:Richtler). The VLT Unit Telescope 4
(Yepun) was used with the FORS2 (FOcal Reducer/low dispersion
Spectrograph) instrument equipped with the Mask EXchange Unit
(MXU). 

The standard resolution collimator used for this program provided a
field-of-view of $6\farcm 8 \times 6\farcm 8$.  

%In the MXU mode, up to ten masks can be loaded into the magazine of
%the Mask EXchange Unit (MXU) during daytime. The selected mask is
%moved in and out of the focal plane by a drive. 

%For objects located close to the edge of the field-of-view (in
%dispersion direction), a part of the spectrum will not be projected
%onto the CCD: for slits near the left (right) border, the blue (red)
%part of the spectrum will be truncated. The mask layout is done by the
%observers prior to observations. This crucial step is described in the
%following section. DISPERSION!

The detector system consisted of two $4096\times2048$ red optimized
CCDs with a pixel size of $15\mu\textrm{m}$. The grism 600B gave a spectral
resolution of about 3 \AA. The spectral coverage was dependent on the slit position
on the mask. Normally, the usable coverage was about 2000 \AA\ with limits on the red
side varying between 5500 and 6500 \AA.
We exposed 8 spectroscopic masks, whose preparation is described in the next section.
Flat fielding was done with internal flat lamps. A He-Ar lamp was used for wavelength calibration. 

The observations are summarized in Table \ref{ov}. 
%Each mask observation was
%divided into two exposures  of 1350 s each.
% minutes- with the
%exception of Field 1\_1 for which three science images were
%BINNING.!

\subsection{Mask Preparation}
\label{sect:masks}
%A very important step in the investigation was the preparation of the
%MXU slit masks.
Preimaging  of the 8 fields (see
Fig.~\ref{fig:slits}) was carried out in October 2006. Each field was
observed in the R filter for 60 seconds. The candidate
selection was based upon the  photometry  in the Washington system 
(Paper I). However, at the time of the mask design, only a preliminary
version of the photometry was available.  Cluster candidates had to fulfill the following
criteria: the allowed color range was $0.9 < C-R < 2.1$, and the
candidates should exhibit a star-like appearance on the pre-images to
distinguish them from background galaxies.  The colour interval has been defined
before we became aware that NGC 1316 hosts many bluer (and younger) clusters (Paper I). 
We moreover avoided  objects brighter than  R=20 mag, only a few bright objects entered 
the sample in an effort to fill the mask.
 
The ESO FORS Instrumental Mask Simulator (FIMS)
software\footnote{available from
  http://www.eso.org/observing/p2pp/OSS/FIMS/FIMS-tool.html} was then
used to select the positions, widths and lengths of the slits. A slit
width of 1\arcsec\  was chosen which tolerates also slightly worse seeing
conditions. 

The choice of the slit lengths was determined by the fact 
% however, was not an easy task. Given
that most targets are very faint and therefore
%one would assume that
 the best
strategy is to measure sky and object in the same
% say  5\arcsec\,
slit. However, this severely constrains the  number of observable
objects per mask, especially in the more crowded  fields. 
But in contrast to previous work, where we wanted to maximize the number
of objects (compare \citealt{richtler04,schuberth06}) we now give more weight to
% Thus,
%another strategy was adopted: Objects and sky positions were observed
%through separate slits of 2\arcsec\, length. Obviously, 
the quality of
the sky subtraction and choose relatively long slits of typically 5\arcsec\ .
 %then depends on the accuracy of the wavelength
%calibration. As it will be shown, it turned out to be satisfactory to
%obtain radial velocities with the desired accuracy. However, work on
%spectral line strengths might be severely hampered by this
%approach.
 After the positioning of the slits for the  selected GC
candidates, the remaining space on the masks (especially in the outer
fields) was used to include additional objects.  Thus, also some background
galaxies and point sources not matching the above mentioned criteria
were observed. 

\subsection{The dataset}
%The observations are summarized in Table \ref{ov}.
 To prevent a severe
contamination from cosmic ray hits, the observation of each mask was
divided into two exposures of 45 min each
%(CHEC THIS
- with the
exception of Field 3 for which three science images were
obtained. In all spectra, the night sky emission lines red-wards of
about 5200\,\AA\, are  by far the most prominent features, i.e. the
spectra of the GC candidates are  sky dominated.  

In addition to the spectroscopic observations, calibration
measurements were obtained during day time.

\begin{figure}[t]
%\centering
\centerline{\resizebox{\hsize}{!}{\includegraphics[angle=0]{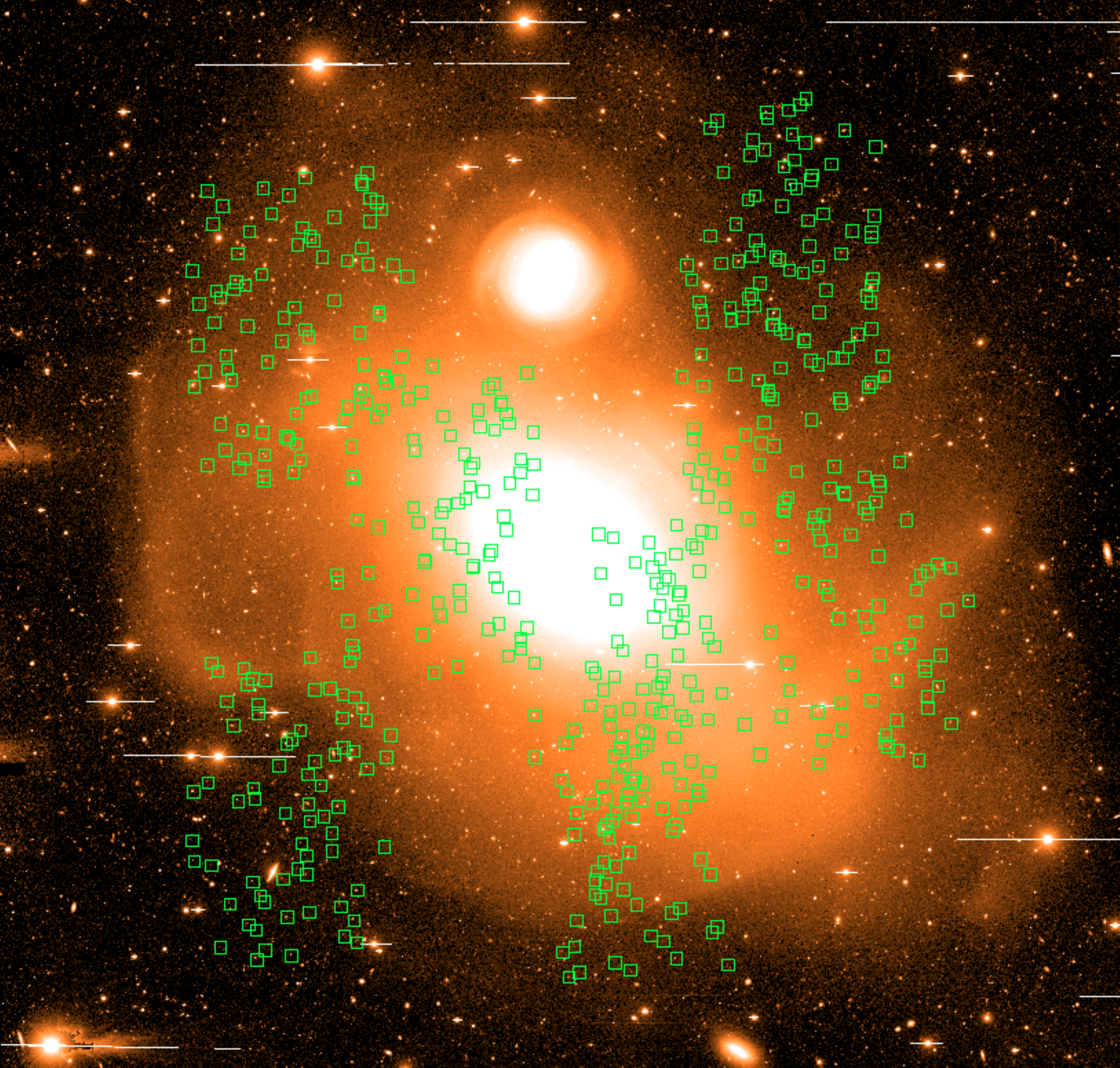}}} 
\caption[Slit positions ]{{Positions of the slits  overlaid on
    a $36\arcmin \times 36\arcmin$ image taken with the MOSAICII camera at the
    4m-Blanco telescope, Cerro Tololo (see Paper I for more details). North is up, east
    to the left.
     }}

\label{fig:slits}
\end{figure}

Fig. \ref{fig:slits} shows the distribution of 562 slits, located on 8 FORS2 fields. Only a minor fraction of these
slits finally provided radial velocities of GCs.

In total, we determined velocities for 177 GCs and 81 stars  (the velocity gap between stars and GCs is sufficiently large for safe classifications).  Five GCs in our sample have been already measured by \citet{goudfrooij01b}. We found  16  quasars, and 117 galaxies.  We did not attempt to derive redshifts for all objects. The remaining
spectra could not be used due to low S/N. 

%\begin{description}
%\item[bias:]
%\item[flat fields:]
%\item[wavelength calibrations:]
%\end{description}

\begin{table*}
\centering
\begin{tabular}{cllrcrcrr}\hline
\hline
{Field} & \multicolumn{2}{c}{Center Position} &
{Exp.~Time}&{Seeing}&{\#\,Slits}&{OB Id}&{Night}&{UT} \\
{} & \multicolumn{2}{c}{{({J\,2000})}} & {{({sec})}} & & & & &{(start)\,} \\
\hline
%\hline
1& 3:23:13.0  & -37:07:20 & 2700  & 0\farcs 8 & 82 & 258629  & 2006-11-15 &  6:49 \\
1& 3:23:13.0  & -37:07:20 & 2700  & 0\farcs 8 & 82 & 258627  & 2006-11-15 &  4:53 \\

%1\_2& 12:42:45.8& 02:39:24.2  & 4500  & 1 \farcs 0&98 &61  &05&07& 01&49 \\
%1\_3& 12:42:50.8& 02:39:21.3 & 5400  & 1 \farcs 2& 106&65  &05&07&
%03&17  \rule[-1.5ex]{0pt}{2ex} \\
%\hline
2& 3:22:58.0  & -37:11:40 & 2700 & 1\farcs 2 & 83 & 258618  & 2006-11-18 &  6:45 \\
2& 3:22:58.0  & -37:11:40 & 2700 & 1\farcs 3 & 83 & 258616  & 2006-11-18 &  7:31 \\
%2\_1& 12:42:17.8& 02:37:27.9& 3600 & 0 \farcs 9 & 101& 51  & 05&07&05&11 \\
%2\_2& 12:42:23.0& 02:37:27.6 & 3600 & 0 \farcs 9& 99& 48  &
%05&08&23&32  \rule[-1.5ex]{0pt}{2ex} \\
%\hline
3& 3:23:13.0  & -37:18:13 & 2700 & 1\farcs 3 & 70 & 258615  & 2006-11-16 &  5:04 \\
3& 3:23:13.0  & -37:18:13 & 2700 & 1\farcs 0 & 70 & 258613  & 2006-11-16 &  5:59 \\
3& 3:23:13.0  & -37:18:13 & 2700 & 1\farcs 3 & 70 & 258615  & 2006-11-30 &  6:22 \\

%3\_1& 12:42:34.2& 02:46:22.4 & 3600 & 0 \farcs 8&113 &77 & 05&08 &01&21 \\
%3\_2& 12:42:28.9& 02:46:20.7 & 3600 & 0 \farcs 6&113 &54 & 05&08&02&31
%\rule[-1.5ex]{0pt}{2ex} \\
%\hline
%4\_1& 12:42:27.1& 02:53:27.4 & 3600 & 1 \farcs 2&98 &44 & 05&10&23&37
%\rule[-1.5ex]{0pt}{2ex} \\
4& 3:22:33.0  & -37:18:39 & 2700 & 1\farcs 2 & 64 & 258630  & 2006-11-19 &  5:30 \\
4& 3:22:33.0  & -37:18:39 & 2700 & 1\farcs 5 & 64 & 258632  & 2006-11-19 &  6:20 \\
%\hline
5& 3:22:32.0  & -37:15:20 & 2700 & 0\farcs 9 & 68 & 258633  & 2006-11-19 &  6:53 \\
5& 3:22:32.0  & -37:15:20 & 2700 & 0\farcs 9 & 68 & 258625  & 2006-11-20 &  4:42 \\ 

%5\_1& 12:43:00.6& 02:46:50.3 & 3600 & 0 \farcs 7 &105 & 59&05&08&
%03&43  \rule[-1.5ex]{0pt}{2ex} \\
6& 3:22:09.0  & -37:13:52 & 2700 & 0\farcs 8 & 66 & 258622  & 2006-11-18 &  5:17 \\
6& 3:22:09.0  & -37:13:52 & 2700 & 0\farcs 8 & 66 & 258624  & 2006-11-18 &  4:21 \\

%6\_1& 12:43:15.0& 02:39:33.7& 3600 & 1 \farcs 2& 107 &53 & 05&10&01&04 \\
%6\_2& 12:43:12.6& 02:39:32.0& 3600 & 0 \farcs 8& 110 & 55 &
%05&09&00&18  \rule[-1.5ex]{0pt}{2ex} \\
%\hline

7& 3:22:20.0  & -37:08:45 & 2700 & 0\farcs 9 & 64 & 258621  & 2006-11-20 &  5:47 \\
7& 3:22:20.0  & -37:08:45 & 2700 & 1\farcs 2 & 64 & 258619  & 2006-11-20 &  7:02 \\

%7\_1& 12:43:33.9& 02:32:12.1 & 3600 & 0 \farcs 9 &109 & 52& 05&09&01&31 \\
%7\_2& 12:43:29.3& 02:32:09.7& 3600 &  0 \farcs 6 &113 & 59& 05&09&02&44 \\
8& 3:22:16.0  & -37:05:36 & 2700 & 0\farcs 8 & 65 & 258610  & 2006-12-21 &  3:07\\
8& 3:22:16.0  & -37:05:36 & 2700 & 1\farcs2  & 65 & 258612 & 2006-12-21 &  2:10 \\
\hline
\hline

\end{tabular}
\caption[Summary of observations]{{Summary of observations (ESO
    program ID 78.B-0856(A)). The seeing values are those recorded by
    the ESO seeing monitor.}}

\label{ov}
\end{table*}

\subsection{Remarks on the reductions and velocity measurements} 
The   reduction procedure and the measurement of radial velocities  have been already described in numerous
other papers, e.g. \citet{schuberth06,richtler04,schuberth10,schuberth12}, so that we can be short.

For basic reduction, spectrum extraction and  wave-length calibration, we used the IRAF-task {\it identify} and {\it apall}. 
%In total, we extracted and calibrated  about 600 spectra, of which 178 were GCs, ?? stars, 16 quasars, and ?? galaxies. 

The radial velocities have been determined
using the cross-correlation IRAF-task {\it fxcor}.  Due to the very different appearance and S/N of the spectra, it turned out to
be impossible to establish a standard procedure, which would always use the same task parameters. Regarding the cross-correlation
interval, we made good experience with the range 4700\AA -5400 \AA.  Clearly defined correlation peaks are connected with
uncertainties around 20-30 km/s.   In the case of faint sources, more than one peak might appear, depending on the exact wavelength
interval, within which the correlation is done. In these cases, we tried out what peak is the most stable against variations of the cross-correlation interval.  The uncertainty then may not be the uncertainty suggested by the
broadness of the correlation peak. We used as templates a high S/N spectrum of NGC 1396, obtained with the same instrumentation during an earlier run \citep{richtler04} and a spectrum of one of the brightest globular clusters in NGC 4636 \citep{schuberth10}  (identification f12-24).

The globular cluster data  are  presented in  Appendix B.
% \citep{kumar12}.
      
\subsection{Comparison with previous measurements}      
\label{sec:previous}
Because the fields  were not strongly  overlapping,  there are only six double measurements, i.e. the same object on two different masks.
In Table \ref{tab:gcsample_points}, they have the identifications gc01214, 2891, 2977, 3151, 4128, 4138. The standard deviation of the velocity differences
is 30 km/s. The small sample size probably prohibits to see more in this value than a rough approximation. 
% which has been used in previous work 
%for estimating the velocity uncertainties
  However, our experience from previous work   \citep{schuberth10,schuberth12} is that  the uncertainties given by {\it fxcor} 
normally are a good approximation of the true uncertainties. 

\citet{goudfrooij01b} measured radial velocities for a small sample of GCs. Their objects are strongly concentrated to the inner regions, so that  we
have only 5 objects in common. Table \ref{tab:goud} shows the common GCs. The  zeropoints agree extremely well, leaving the velocities
of \citet{goudfrooij01b} by only a mean of 5.4 km/s higher than our velocities. This agreement can be partly coincidental, but at least it shows that
the two velocity samples do not differ greatly in their zeropoints. 

\begin{table}[h]
\caption{Comparison of the common GCs in the sample of \citet{goudfrooij01b} and the present sample. The columns are: Identifier and velocity of Goudfrooij et al.,
velocity as in the present paper, identifier in Tables \ref{tab:gcsample_points} and \ref{tab:gcsample}.}
\begin{center}
\begin{tabular}{cccc}

\hline
$ID_{G}$ & $v_r (Goud)$ &  $v_r (Ri)$ & $ ID $ \\
\hline
   123 &  1966$\pm$1 & 1977$\pm10$  &   gc03384  \\
   217 &  1840$\pm$8 & 1855$\pm25$  & gc03318 \\
   121 &  1627$\pm$155 & 1618$\pm27$  & gc08412 \\
   204 &  1992$\pm$19 & 1976$\pm34$ & gc02997\\
   203 &  1639$\pm$35 & 1610$\pm21$ & gc01324\\
\hline
\end{tabular}
\end{center}
\label{tab:goud}
\end{table}%
  
\begin{figure}[]
\begin{center}
\includegraphics[width=0.4\textwidth]{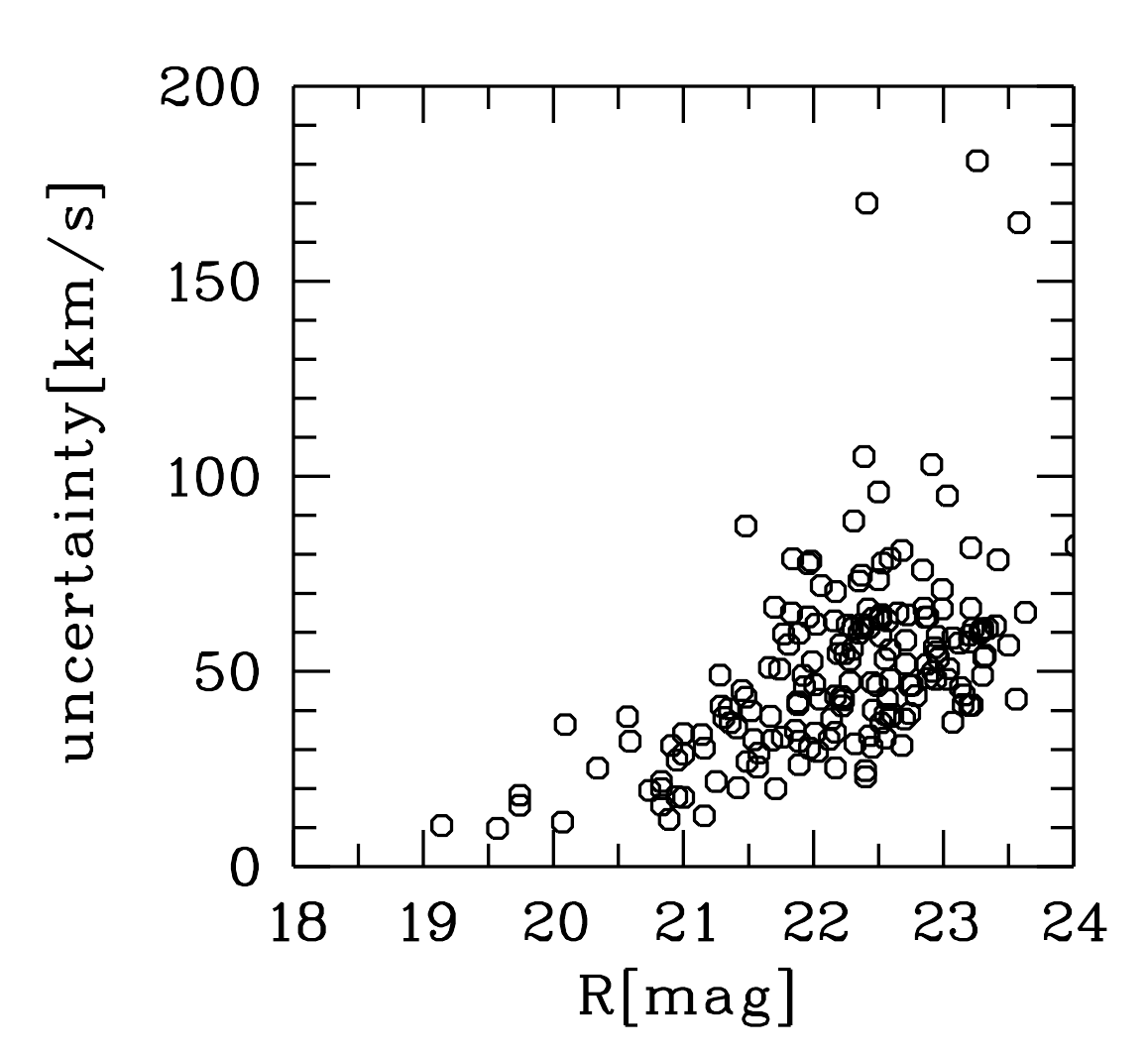}
\caption{Uncertainties of the radial velocities in dependence from the R-magnitude. }
\label{fig:uncertainties}
\end{center}
\end{figure}

Fig.\ref{fig:uncertainties} shows the velocity uncertainties in dependence on the R-magnitude. The uncertainties are directly taken
from {\it fxcor}. They  cluster around 50 km/s as in previous work. The three outliers with errors around 150 km/s are the objects with the photometric identification numbers (Table \ref{tab:gcsample_points}) 341,1278, 3025.
Two of them are very faint (1278,3025),  and the spectrum of 341 might be badly extracted.

\section{Population properties and velocities}

\subsection{Colour-magnitude diagram and colour distribution}

\begin{figure}[]
\begin{center}
\includegraphics[width=0.4\textwidth]{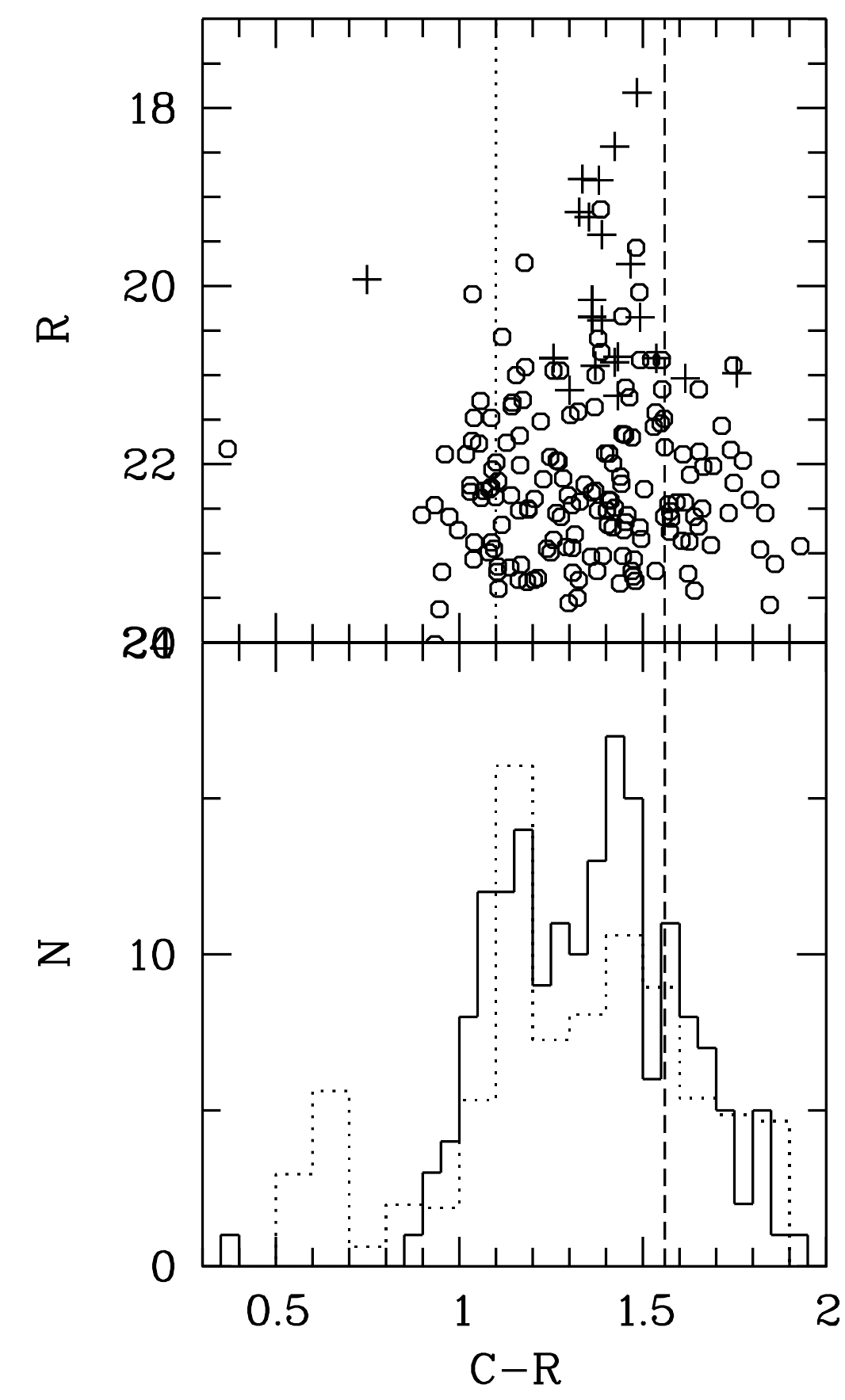}
\caption{Upper panel: the CMD of confirmed globular clusters in NGC 1316 in the Washington system, marked by open circles.
The photometric data are taken from Paper I.  20 objects from \citet{goudfrooij01b} are denoted by crosses. The dashed vertical line denotes the galaxy colour. The dotted vertical marks the colour limit for old, metal-poor globular clusters. Note the excess of
clusters blueward of  this limit which corresponds to ages of about 1 Gyr, if  solar metallicity is assumed. Also note the object at C-R= 0.4. Lower panel: The 
corresponding colour histograms (solid histogram, only for our sample).  The dotted histogram (scaled down for a convenient display) is the more complete photometric sample from Paper I. The two well defined peaks, already indicated in \citet{goudfrooij01a} probably mark two epochs of high star formation rates. }
\label{fig:CMD}
\end{center}
\end{figure}

With the GC colours available as well as radial velocities from the present study,  we show a  ''clean" CMD for confirmed GCs in Fig.\ref{fig:CMD} (upper panel).  This CMD shows the same features, which
have been found already  in Paper I with a larger, but contaminated photometric sample. The dashed vertical line
indicates the galaxy colour, as it has been measured outside the inner dust structure (Paper I).  The dotted vertical line denotes the colour (C-R = 1.06) for clusters with an age of 10 Gyr and a metallicity of z=0.0002 \citep{marigo08}.
Clusters bluer than this must be even more metal-poor (however, colour and metallicity at these low metallicity levels
are not related in a simple manner, see e.g. \citealt{richtler13}) or younger.
%limit for old, metal-poor clusters, which by colours alone cannot be distinguished from younger,
%metal-rich clusters.  
To our sample we add  20 GCs from \citet{goudfrooij01b}, which are coded as crosses. Their Washington
colours have not been measured directly, but were transformed on the  basis of Fig.1 from Paper I, using the B-I colours given by Goudfrooij et al. . 
These objects are strongly concentrated towards small radii, thus individual reddening can be an issue. 
%One striking feature is the  quite small colour interval around C-R=1.4 which is occupied by bright clusters. It may indicate
%that they stem from the same star formation epoch 

 We also find some GCs distinctly bluer than C-R=1.0. These clusters cannot be old GCs. 
 Note the outlying object at C-R=0.4, which  has an age around 0.5 Gyr (metallicity is not anymore a critical parameter at these blue colours). 
%Since the spectroscopic mask design was biased for GCs within the colour range of old GCs, this
%detection is serendipitous.
 As the more complete photometry of Paper I shows, such young objects are rare, but GCs as blue as
C-R=0.8 are common.  
If these objects have their origin in    star formation events  which occurred later than indicated by the peak at C-R=1.4, the assumption is reasonable that
they possess at least solar metallicity.   As reference values we use ages for theoretical Washington colours for single stellar populations, taken
from \citet{marigo08} and graphically displayed in Fig.1 in Paper I. Since we selected our spectroscopic sample with the help of our photometric data, but prior
to the knowledge provided by Paper I, objects bluer than C-R=1.0 were only serendipitously targeted to fill the spectroscopic masks. 

Clusters redder than the galaxy light must be metal-rich and quite old.  The old metal-poor clusters, on the other hand,
cannot be distinguished from younger, more metal-rich clusters, 
%in our sample,
 but, as argued in Paper I, they should not be too many.
% Because the galaxy light is composed from
%a variety of stellar populations,   

The lower panel shows the corresponding colour histogram. The striking two peaks at C-R=1.4 and C-R=1.1 match those which have been photometrically identified. They  are even
weakly indicated in the smaller, but
also clean sample of \citet{goudfrooij01b}.  In Paper I, we show that these  peaks  are a property of the colour distribution
only for bright clusters (R$<$23 mag). They largely vanish if also fainter clusters are included.  Paper I tentatively interprets these peaks
as signatures of starbursts with ages 1.8 Gyr and 0.8 Gyr, respectively. 
 
%\subsection{Sample completeness}

NGC 1316 has an interesting companion galaxy, NGC 1317 (see the appendix). Its systemic velocity is 1941 km/s.  There is no indication for the photometric GC sample (Paper I), 
  that 
GCs from NGC 1317  would be visible in  the system of NGC1316. Moreover, the region  of NGC 1317 is not covered by masks, but we cannot exclude that a few GCs belong to 
NGC 1317. None of the results of this contribution, however, depend on this possibility.

%\section{Velocity distributions}

\subsection{Velocities, colours, magnitudes}

\begin{figure}[]
\begin{center}
\includegraphics[width=0.5\textwidth]{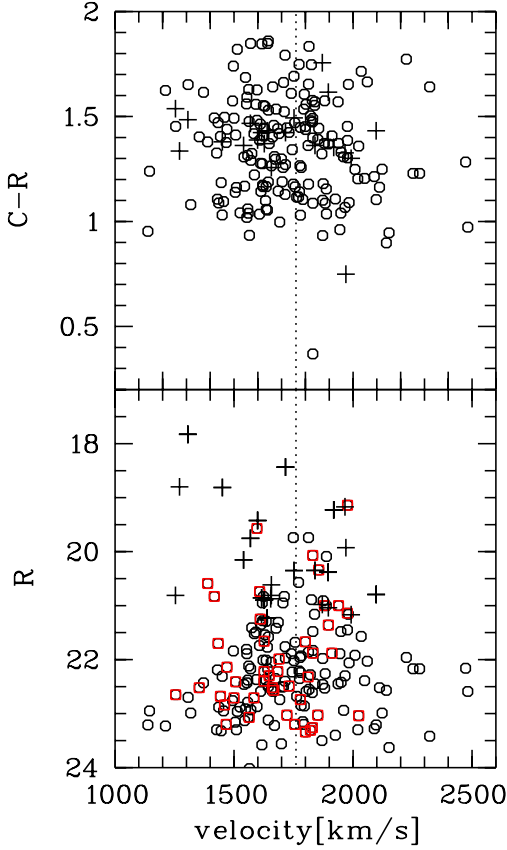}
\caption{Upper panel: velocities versus colour. Open circles denote the present GC sample, crosses the objects of \citet{goudfrooij01b}. In both panels, the systemic velocity is marked by the dotted vertical line. The bimodal colour distribution is well visible. Note that the objects belonging to the peak at C-R=1.4 avoid velocities
higher than the systemic velocity.  Lower panel: velocities versus R-magnitude. The brightest clusters do not show a kinematic affinity to the bulge
population. Note also the increasing velocity
dispersion for clusters fainter than R= 21.5.  
  }
\label{fig:col_vel}
\end{center}
\end{figure}

 A closer look at the relation between velocities, colours, and magnitudes reveals interesting facts.
In the upper panel of Fig.\ref{fig:col_vel}, velocities are plotted versus colours. The double peak structure in the colours is clearly
discernable. A strange pattern is that the peak at C-R=1.4 is populated preferably by objects with radial velocities lower
than the systemic velocity.
%Moreover, high velocities are apparently underrepresented in the peak at C-R=1.4.
% The  selection 1.35 $<$ C-R $<$ 1.5 and

 The lower panel  shows that  clusters  brighter than about R=21.5 {\it avoid the systemic velocity of NGC 1316}.  Objects with colours 1.35$<$ C-R $<$1.5
 (the pronounced peak in Fig.\ref{fig:CMD}) are marked in red.  
 %The velocities of Goudfrooij et al.'s objects are known since a decade, but to
 %our knowledge this had not yet been remarked. 
  Our GCs confirm the trend, which is visible already in the sample of Goudfrooij et al., also for  fainter clusters.  One would expect that, if
 the majority of the bright clusters in red are clusters formed in the starburst with an age of about 2 Gyr (the dominant bulge population),  they would
 also be kinematically connected to the bulge. However,  the velocity field of the bulge  is not known.
  The field stars related to the bright GC population should show the same kinematics,
 %Instead, these objects prefer low velocities, 
 % be   Considering only our sample, 
%distribution of the velocities is very dependent on the brightness: the brightest clusters in our sample  have a smaller dispersion than the 
%fainter clusters. However, the addition of the Goudfrooij et al.-sample shows that  clusters brighter than R=21mag display a broad range of
 which does not fit at all to a Gaussian distribution around the systemic velocity.    Relative  radial velocities as high 
 as 500 km/s and more indicate that
 these clusters are deep in the potential well and that their orbits are elongated. Even if these objects are now projected 
 onto the bulge, it may be that  their place of birth was not the bulge, but  a star burst in
 one of the merger components in an early  stage of the merger.

The second striking observation is that the velocity distribution becomes broader with decreasing brightness. 
In Paper I it is shown that the bimodal colour distribution disappears if fainter clusters are included. It is therefore 
plausible to assume that the bright cluster population consists mainly of intermediate-age clusters belonging to a  population with the complex kinematics of a merger/star burst  situation still preserving,
while the fainter   older clusters, filling a larger volume around   NGC 1316, are
progressively mixed in.

\subsection{Velocity histograms}

\begin{figure*}[]
\begin{center}
\includegraphics[width=0.9\textwidth]{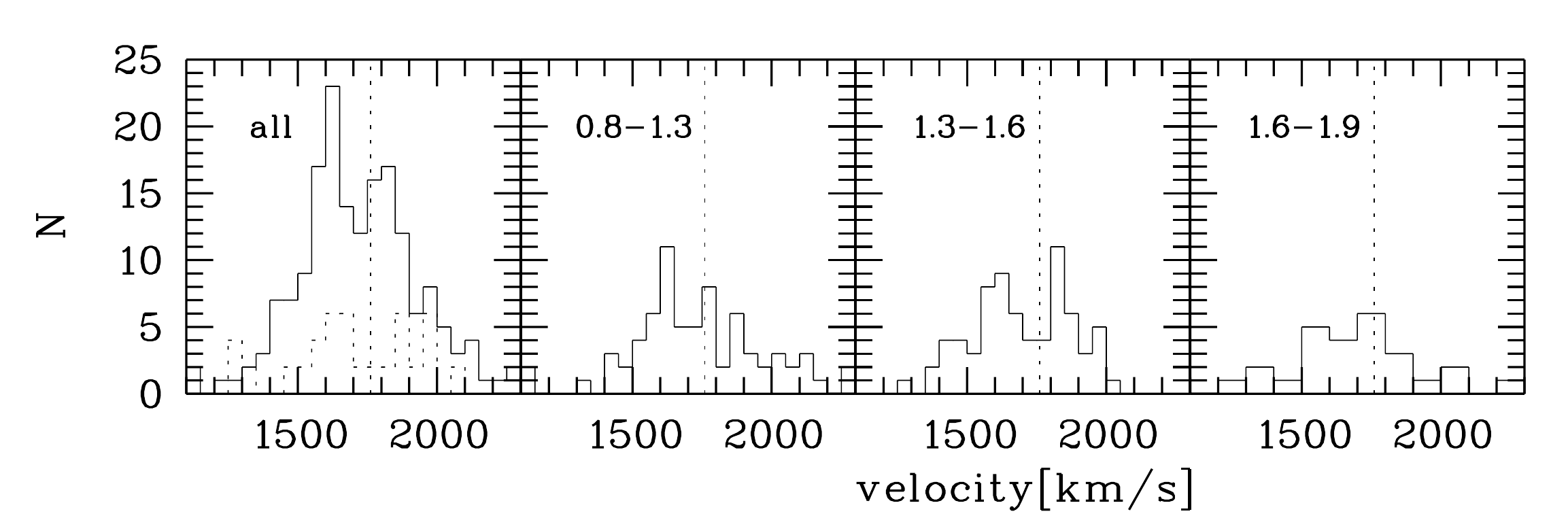}
\caption{Radial velocity histograms for different colour intervals, corresponding to different  cluster populations (see text).
Typical uncertainties are of the order 50-80 km/s (compare Fig.\ref{fig:uncertainties}).
The vertical dotted line indicates the systemic velocity. The dotted histogram in the left panel is the velocity histogram of \citet{goudfrooij01b}, where
the two velocity peaks  already  are discernable. We interpret the peak at 1800 km/s as the expected peak at the systemic velocity. The peak
at 1600 km/s is caused by a dominance of clusters with this velocity in the western part of NGC 1316.}
\label{fig:veldistri}
\end{center}
\end{figure*}

The velocity histograms in several colour bins are shown in Fig.\ref{fig:veldistri}. The colour bins refer to the bins used in Paper I 
to characterize the population mix of globular clusters and which appears as a reasonable binning guided by Fig.\ref{fig:CMD}.
 The interval 0.8 $<$ C-R $<$ 1.3 contains an unknown proportion of old, metal-poor
clusters and clusters younger than about 1 Gyr. The interval 1.3 $<$ C-R $<$ 1.6  contains the bulk of intermediate-age clusters.
In   the interval 1.6 $<$ C-R $<$ 1.9, one expects to find old, metal-rich clusters. Instead of a unimodal Gaussian-like distribution, one sees 
in all histograms, except for the reddest clusters, two
velocity peaks, which are best defined for the bluer clusters. The higher velocity peak agrees well with the systemic velocity of NGC 1316, but
the low velocity peak at 1600 km/s indicates a peculiarity. 

The nature of this peculiarity can perhaps be inferred from Fig.\ref{fig:xyplot_vel}.  The objects  populating the peak are preferentially located on  the western side
of NGC 1316, occupying a large interval of position angles. This is suggestive of a disk-like distribution of GCs seen from face-on.  
%A plausible assumption is that of a disk-like distribution. 

Of course that does not mean that the supposed disk population is present only in the interval 1550-1650. The complete dispersion in z-direction
is unknown and  uncertainties in the velocities widen an intrinsically sharp distribution. 

\subsection{Two-dimensional distribution}
\begin{figure*}[ht!]
\begin{center}
\includegraphics[width=0.7\textwidth]{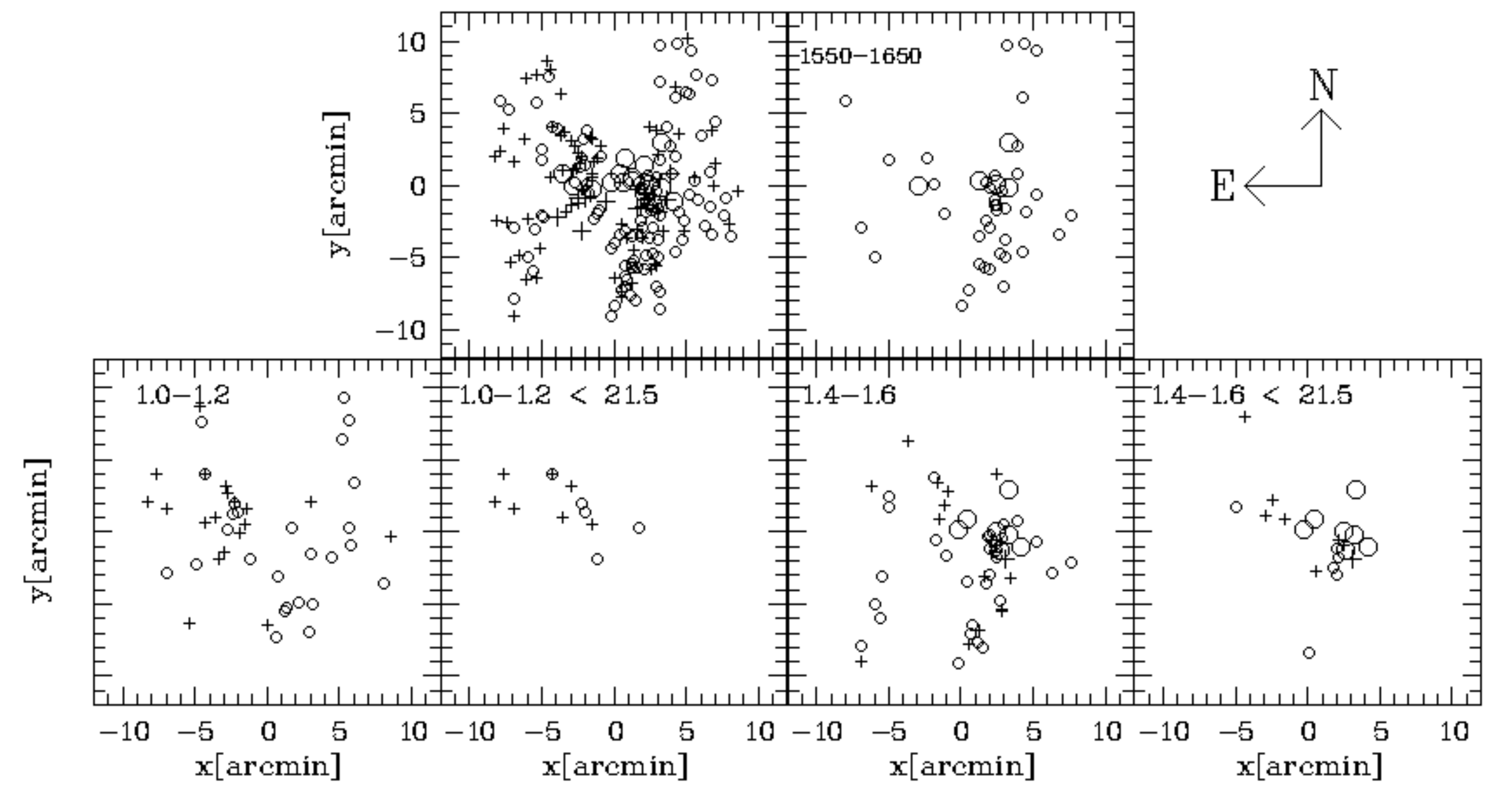}
\caption{Spatial distribution of globular clusters under various selections with the center of NGC 1316 as the origin.  Crosses are clusters
with velocities higher than the systemic velocity, open circles denote lower velocities. Large symbols are objects from the
sample of \citet{goudfrooij01b}. Upper left panel: Full sample, which is cleanly separated in
an eastern and a western part. Upper right panel: clusters with
velocities between 1550 km/s and 1650 km/s. There is an overwhelming dominance of clusters on the western side of NGC 1316, indicating that
these clusters intrinsically have a disk-like distribution, seen almost face-on. 
Lower leftmost panel: clusters in the colour interval 1.0$<$C-R$<$1.2 (blue peak).  Next panel: additional selection with only 
clusters brighter than R=21.5. These are {\it bona fide} younger clusters with ages around 0.8 Gyr. Almost all are located in the north-eastern bulge region.  Next panel: clusters in the colour interval 1.4$<$C-R$<$1.6 (red peak). Lower rightmost panel: additional selection with only
clusters brighter than R=21.5.   These are {\it bona fide} younger clusters with ages around 2 Gyr and appear dominantly in the
bulge region.
% Right panel: velocity distribution of the western part (dotted histogram)
%and the eastern part (solid histogram).
 }
\label{fig:xyplot_vel}
\end{center}
\end{figure*}

Fig.\ref{fig:xyplot_vel} shows the two-dimensional distribution of clusters for several selection of
velocities or colours/magnitudes. Because we could not achieve a complete azimuthal
coverage,  we have well defined western and eastern parts. Crosses denote velocities higher than the
systemic velocity of 1760 km/s, circles lower velocities. North is up, east to the left.

The upper left panel contains the full sample.
The larger symbols are objects from the HST cluster sample of \citet{goudfrooij01b},
% concentrated to the innermost
%region.  
 %The innermost region is
which   populate the innermost region,
% the cluster sample of \citet{goudfrooij01b},
%observed with HST, 
where we do not have objects due to the bright galaxy light. A difference between the 
western and eastern part is not discernable. This changes dramatically, if we select only the peak at 1600 km/s,
which is the upper right panel with the selection indicated.  
 These objects dominantly populate the western part and it is tempting to imagine 
that we are looking onto a large disk or at least a structure which is thin along the line-of-sight. A few of these
objects may belong to Schweizer's   L1-structure \citep{schweizer81,richtler13} (see Fig.\ref{fig:1316sub}), but the largest concentration is found still within the morphological
bulge.

The lower panels show selections according to colour and magnitude. The left panel selects the interval
1.0 $<$C-R$<$1.2, which may contain younger clusters, but also old, metal-poor objects. Here we observe
that the majority of objects with velocities higher than the systemic velocity are located on the eastern part.
The following panel selects from this sample only objects brighter than R=21.5 mag. These bright clusters
have a higher probability to be young (about 0.8 Gyr) than fainter objects. Practically all clusters are on the 
north-eastern side. The next panel selects 1.4$<$C-R$<$1.6, an interval which hosts the brightest clusters of intermediate
age.   Except for the concentration at the south-western side, there is nothing striking. Selecting only
the brightest clusters (which with the highest probability are clusters of age about 2 Gyr), one recognizes 
the major axis of NGC 1316. 

Bright clusters therefore are probably connected with the bulge (we have no complete knowledge
about the occurrence of bright cluster at large radii), but the population is inhomogeneous regarding
colour and spatial distribution.

%However, we do not see any preferred population.

\subsection{Velocities and radial velocity dispersions}

\begin{figure}[h]
\begin{center}
\includegraphics[width=0.4\textwidth]{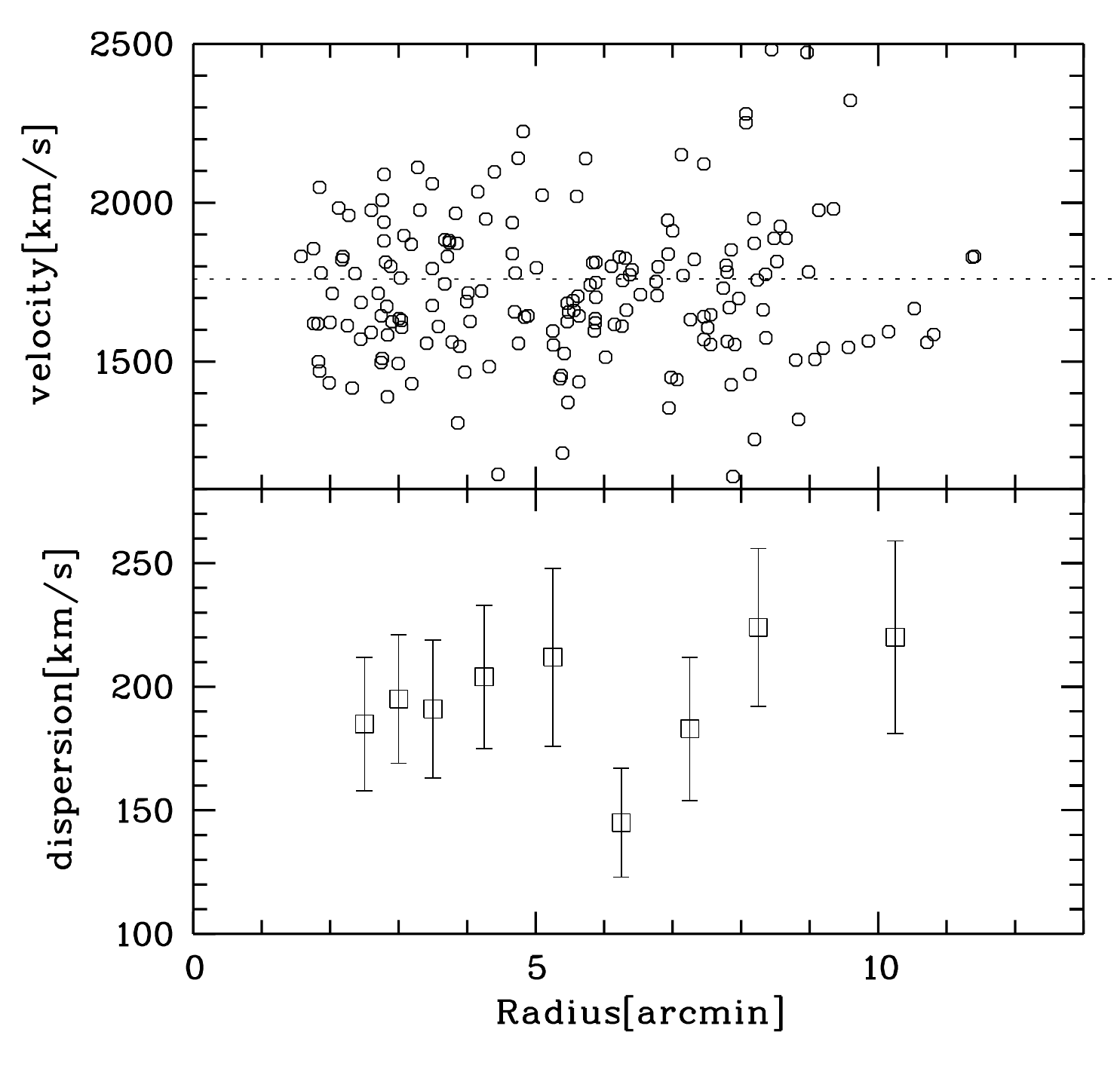}
\caption{Upper panel: Radial velocities vs. galactocentric distance. Lower panel: Velocity dispersions for slightly overlapping
radial bins. }
\label{fig:disp}
\end{center}
\end{figure}

In the following, we consider only our sample and disregard the objects of \citet{goudfrooij01b}.  
Fig.\ref{fig:disp} shows in its upper panel the radial velocities versus the radial distance, in its lower panel the velocity dispersions
in slightly overlapping bins. The bin widths of 1\arcmin\ and 1.5\arcmin\ have been chosen to contain about 30 objects each in order
to get a statistically meaningful velocity dispersion value.   The velocity dispersions have been determined using the dispersion estimator of \citet{pryor93}, adopting a systemic velocity of 1760 km/s. The uncertainties have been evaluated according to the quoted maximum-likelihood formalism
using the individual uncertainties
of the objects. Since the bins are not independent, the error bars probably overestimate the true uncertainty. However, the physical meaning of the velocity
dispersion is not interpretable straightforwardly. It has a well known dynamical meaning for example  in case of a non-rotating spherical or elliptical system in equilibrium, while
the underlying symmetry in NGC 1316 is elliptical only in the inner region. It is clear without any test for Gaussianity that the dispersions for  radii 
larger than 5\arcmin\ do not represent the dispersion of a Gaussian velocity distribution. To what extent the dominance
of velocities smaller than the systemic velocity can be understood as a sample effect, is difficult to evaluate, but outside the
bulge are
simply more clusters in the S-W-region, and it is this region which contributes with a sharp peak at about 1600 km/s. For radii smaller than 5\arcmin\ (which is the bulge) it is not so obvious.
A Wilkinson-Shapiro test gives a p-value of 0.076, so the hypothesis that these objects follow a Gaussian distribution, is statistically valid, but not probable from other considerations. We come back to that shortly.

The radial increase of the dispersion probably is real, but, as Fig.\ref{fig:col_vel} suggests, it may be caused by the bias towards bright
clusters for smaller radii. These bright clusters  show  a smaller  dispersion. Moreover,  the relative contribution of
old metal-poor clusters might be higher for larger radii, which we expect to increase the line-of-sight dispersion.
   
Table \ref{tab:dispersions} lists the values in  Fig.\ref{fig:disp}.  It also lists the dispersion values for a subsample fainter than R=21.5, which
is more appropriate for being compared to a spherical model than the full sample (see Sect.\ref{sec:dynamics}), although the difference is
hardly noticeable. Moreover, dispersions for an inner and an outer subsample as well as for two colour selections are given. The dispersions
obviously depend in an irregular manner  on the binning and sampling. The most natural assumption is that of a radially constant velocity dispersion.  
   
%  \begin{table}[h!]
%\caption{Velocity dispersions for different radial and magnitude samples. }
%\begin{center}
%\resizebox{9cm}{!}{ 
%\begin{tabular}{ccccccc}

%  all & $<$ 4', blue & $<$ 4',interm. & $<$4',red & $>$4', blue & $>$4', interm. &$>$4', red\\
%\hline
 %  201$\pm$12 (172) & 177 $\pm$ 28 & 181 $\pm$ 26 & only 7& 217$\pm$24 (45)& 178$\pm$28 (39)  & 229 $\pm$ 35 (25)\\
% \hline
% \hline F
 
%    &  $<$4', $>$22 mag&    $>$4', $<$22 mag& $<$4', $<$ 22   &  &  &  \\
    %\hline 
%%
%  \end{tabular}
 %}
%\end{center}
%\tablefoot{Bin widths are 0.1 mag. The background is defined by r$>$13\arcmin. The uncertainties are based on the
%square root of the raw counts. Negative values are kept for formal correctness.}
%\label{tab:dispersions}
%\end{table}%

\begin{table}[h!]
\caption{Velocity dispersions for different radial and magnitude/colour samples. Sample sizes are given in
parentheses. Dispersion values for the fainter sample are used in Fig.\ref{fig:models}.}
\begin{center}
\resizebox{9cm}{!}{ 
\begin{tabular}{cccc}
\hline
Radius[arcmin] &    $\sigma$[km/s]  & Radius[arcmin]    &$\sigma$[km/s] \\
\hline
                          &   no selections      &       &   R$>$ 21.5  \\
\hline
2-3           &  185$\pm$27 (27) &   2-3.5    &  185$\pm$28  (26)   \\
2.5 - 3.5   &   195$\pm$26 (30)         &  3-4.5    & 208$\pm$31 (25)     \\
3 - 4         &   191$\pm$28 (26)        &    4-5.5  &  241$\pm$37 (26) \\
3.5 - 5         &   204$\pm$28 (29)        &   5-6.5    & 199$\pm$35 (32) \\
4.5 - 6         &   212$\pm$29 (36)        &    6-7.5  &  189$\pm$32 (33) \\
5.5 - 7         &   145$\pm$22 (33)        &     7-8.5 & 236$\pm$33 (28)    \\
6.5 - 8         &   183$\pm$29 (39)        &    -   &      - \\
7.5 - 9         &   224$\pm$32 (29)        &    -    &     - \\
8.5 - 12         &   220$\pm$39 (19)        &   -    &    -   \\
\hline
                 &                                         &                 1.0 $< $C-R$ <$ 1.2  &  1.4 $< $C-R$ <$ 1.6\\ 
\hline
all                   &   201$\pm$12 (175)                      & 182$\pm$25 (45)    &     174$\pm$26 (49) \\
$<$5             &      194$\pm$16 (79)                       &              \\
$>$5            &    206$\pm$17  (96)                            &              \\
\hline
\end{tabular}
 }
\end{center}
\label{tab:dispersions}
\end{table}

\subsection{Radial distribution of clusters}
In paper I we showed that the radial distribution of cluster candidates is quite different for different colour intervals. Particularly
the distribution of cluster candidates with  intermediate colours does not follow a uniform power-law, which means that the 
three-dimensionable distribution is strongly substructured.

On the other hand, we want to use the GC kinematics   in the frame of a spherical model to constrain the potential. The total
sample is apparently not suitable, as the kinematics  of the bright clusters show. Being led by Fig.\ref{fig:col_vel}, we therefore choose all clusters fainter than $m_R = 21.5$ to define within our possibilities a sample which is the best approximation to a spherically homogeneous sample in the
sense that no strong association to the inner bulge population, like the case of GC candidates of intermediate colour, is visible. 

\begin{figure}[h]
\begin{center}
\includegraphics[width=0.4\textwidth]{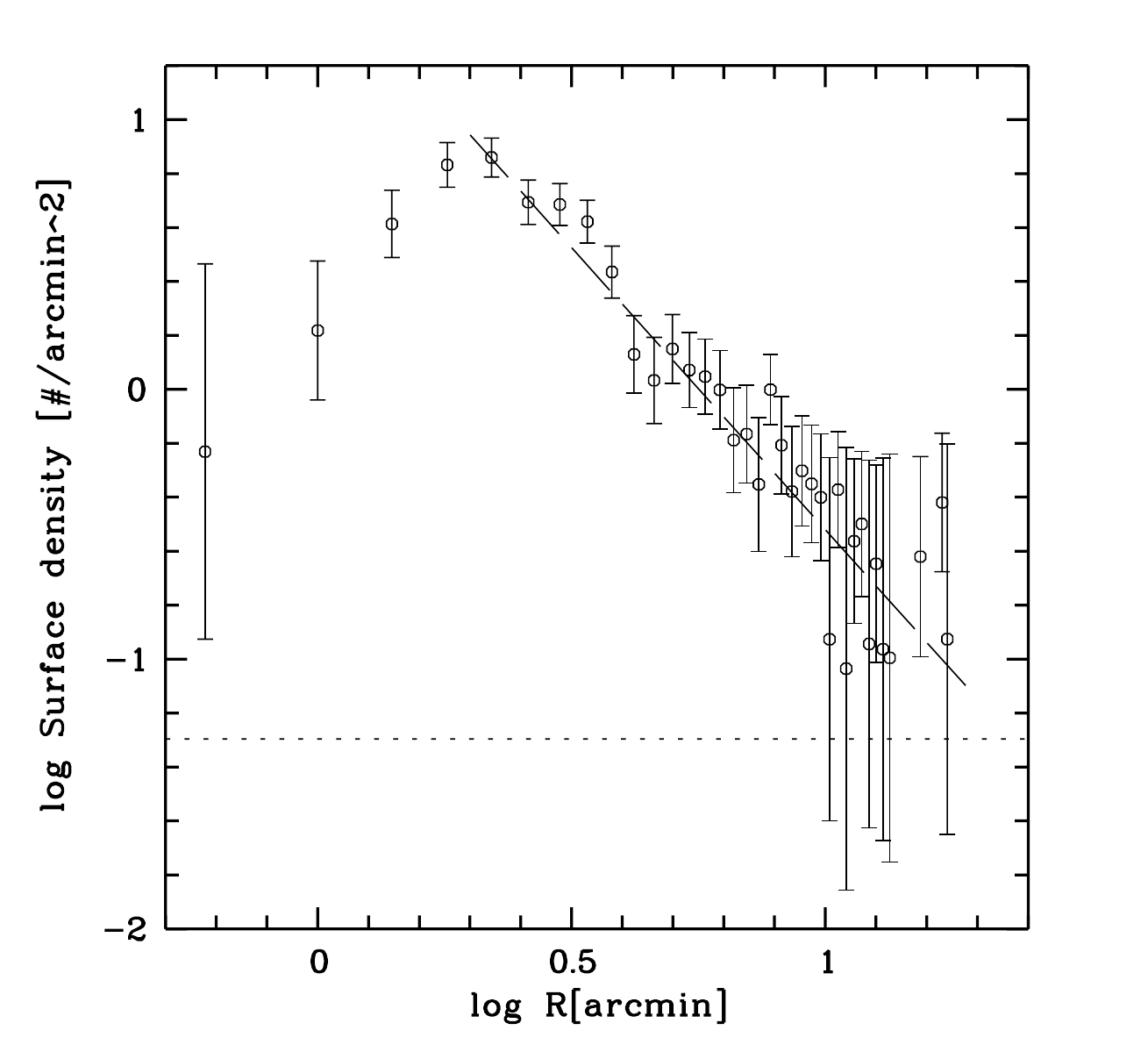}
\caption{The radial density profile for GC candidates fainter than $m_R = 22$ mag derived with the database of Paper I.  The selection of point sources has been
done as described in Paper I. The horizontal dashed line denotes the background density of GC candidates.  See equation \ref{eq:profil} for
more details.}
\label{fig:radprofile}
\end{center}
\end{figure}

For evaluating the number density profile, we use the photometric database from Paper I (the point-source catalog is available on-line) and select
point-sources in the magnitude interval $22 < m_R < 24$.
Fig.\ref{fig:radprofile} shows the resulting  surface densities. The counts in the inner region become severely incomplete  due to the galaxy
brightness, but  from 2\arcmin\ outwards, the counts are fairly complete. The horizontal dotted line marks the background outside of 13\arcmin\ .
The long-dashed  line represents the "beta-model"
\begin{equation}
n(r)  = 2100 \cdot (1+(r/r_c)^2)^{-1}
\label{eq:profil}
\end{equation}
with n(r) as the surface density in numbers/square arcmin and $r_c$= 8\arcsec\ as a scale radius.  Except for the factor, which has been fitted, this is the spherical model for
the galaxy light from paper I. 
%It is re-assuring that the GC distribution can be  modelled by the galaxy light.
 This is a clear difference to giant elliptical
galaxies, where at least the metal-poor GC subpopulation shows a shallower profile than the galaxy light.    
 This is normally interpreted as a result  of the accretion of dwarf galaxies (e.g. \citealt{richtler13}) donating metal-poor clusters. Fig.11 in Paper I  indeed shows a somewhat shallower profile
for the blue clusters. But in NGC 1316, the blue clusters are a mix with unknown fractions of old, metal-poor and younger clusters.  As the 
intermediate clusters from Fig.11 in Paper I demonstrate, a shallow density profile is probably caused by younger clusters.  The fraction of old, metal-poor 
objects in  our faint cluster sample is unknown,  but it is plausible that their number density profile is not significantly different  from old, metal-rich clusters,
because the pre-merger populations are expected to be dynamically well mixed.

\section{Rotation  of  galaxy and clusters}
The bulge of NGC 1316 rotates along its major axis \citep{donofrio95, arnaboldi98,bedregal06} with an amplitude of about 100 km/s. 
Long-slit observations showed this rotation signal out to a radius of about 2.5\arcmin\ \citep{bedregal06}.
%Since a 2-dimensional velocity field of the stellar light is not yet available, it was not yet known whether the position angle of the maximum
%rotation amplitude coincides with the major axis.
The kinematics of globular clusters might be related to the kinematics
of the stellar population, particularly for clusters of intermediate colour which show the strongest link with the bulge (Paper I). 
Moreover, our data provide the possibility to enlarge the radius of measured rotation and  determine the kinematical axis of the galaxy light.

\subsection{Galaxy velocities and rotation signal}

 Some of our
targets are so close to the galaxy's center that their ''sky'' spectra can be used to measure the radial velocity of the galaxy light at the
location of the target. Since the masks were not designed for this purpose, we can use only slits where corresponding skyslits (now the
real sky) can be found. In practice, we constructed average skyslits from regions of the mask with low background intensity and subtracted
them from a given target background slit. Offsets along the dispersion direction obviously cause systematic errors, since the shape of
the spectra without flux calibration depends on the position within the mask. Finally, we selected 72 slits from two masks, where we measured
the radial velocity by cross-correlation with  NGC 1396 as the template.
% (see also \cite{schuberth10}).

\begin{figure}[h]
\begin{center}
\includegraphics[width=0.5\textwidth]{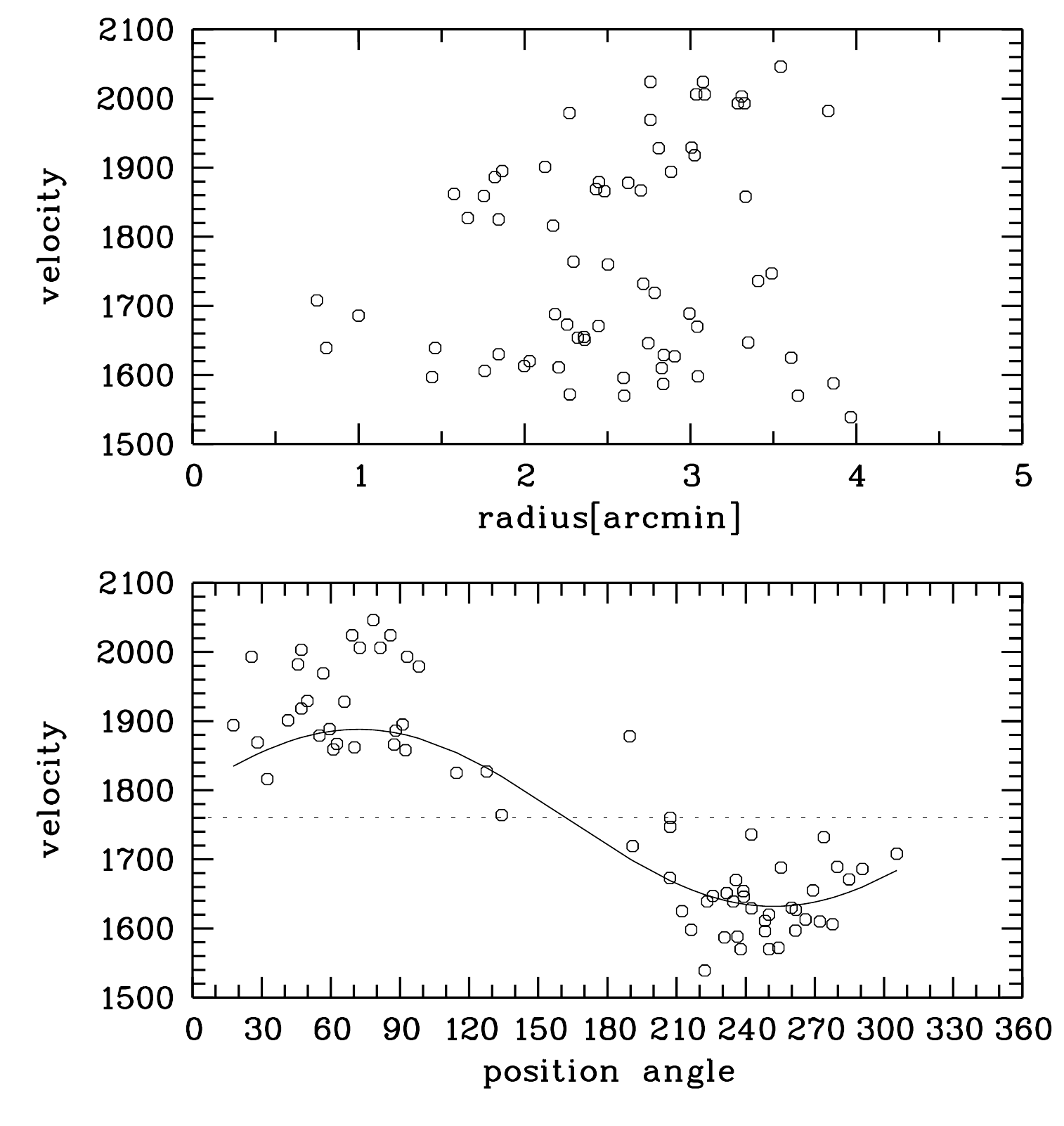}
\caption{Upper panel: the radial distribution of galaxy velocities. Lower panel: 
Distribution of galaxy velocities over position angles. The displayed rotation signal (represented by a sine-curve) has an amplitude of 128  km/s, corresponding to
the inner clusters, and peaks
at 72$^\circ$.
% The amplitude is slightly higher than that of Bedregal et al. 
The centre of rotation, indicated by the horizontal line, has the value
1760 km/s.
  }
\label{fig:galvel}
\end{center}
\end{figure}

Fig.\ref{fig:galvel} shows the resulting velocities  in dependence of radius (upper panel) and position angle (lower panel). We fit the rotation signal by a sine
$v_{rot} = A \sin(\phi + \phi_0) + v_0$, where A is the amplitude, $\phi_0$  the phase constant, and $v_0$ the radial velocity of the
center of rotation which one identifies with the velocity of NGC 1316.  In the upper panel, it is striking that the velocities higher than 1760 km/s
seem to increase with radius, while the lower velocities are more or less constant. The larger velocity scatter in the two regimes of position angle
do not reflect the errors, but should be real, indicating than the rotation cannot be characterized by one amplitude only.    This is strengthened
by the comparison with planetary nebulae. See more remarks in Section \ref{sec:pne}. 

\begin{figure}[h]
\begin{center}
\includegraphics[width=0.5\textwidth]{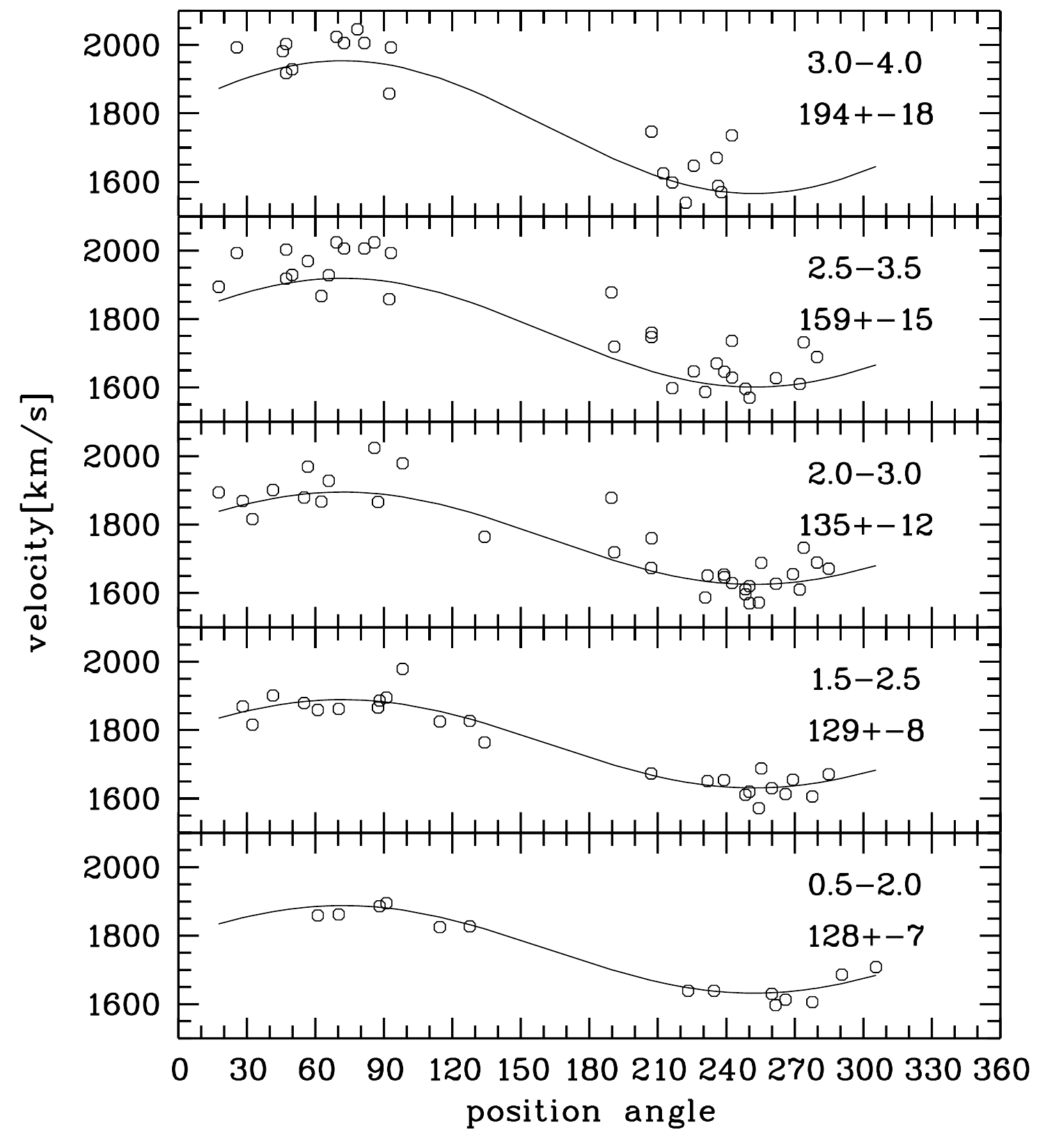}
\caption{Rotation signature of the galaxy light in different radial bins. The bin is indicated in arcmin, the rotation amplitude in km/s. The phase and the center of rotation has been kept at fixed values
of 18$^\circ$ and 1760 km/s, respectively, which are excellent representations for the three innermost bins.  The two outer bins start to deviate to higher values, perhaps indicating a progressive
deviation from a well defined rotation. }
\label{fig:galrot}
\end{center}
\end{figure}

The resulting values of A for the entire sample and radial subsamples
are given in Fig. \ref{fig:galrot}. Since the subsamples do not differ significantly regarding the values of $\phi_0$ and $v_0$, we fix
them to $\phi_0$=18$^\circ$, the value for the innermost sample, and $v_0$=1760 km/s, the systemic velocity.  Thus a position angle of 72$^\circ$ marks the major axis of the rotation. Interestingly,
 this angle is larger by at least  10$^\circ$ than the position angle of the optical major axis, for which \citep{schweizer80} quotes a position angle of 50$^\circ$ at 1.0\arcmin\ and 60$^\circ$ at 2.5\arcmin. 
 
  % which may indicate that the velocity field of the
% galaxy is more complex than suggested by a simple elliptical geometry. 

%This difference has been called T
 About 10\% of the ATLAS$^{3D}$-galaxies show this "misalignment angle" (\citealt{statler91,krajnovic11}).
 % see more remarks in Section \ref{sec:rotdiscussion}).
 Out to a galactocentric
distance of 2.5\arcmin\ (15.5kpc),  the amplitude of the rotation signal is practically constant and very well defined. The outermost two bins are 
somewhat elevated, but whether this is due
to an intrinsically higher rotation or due to a more complex velocity field, cannot be decided. 
The comparison with \citet{bedregal06} shows a very good agreement in the region of overlap. 
%and we conclude that out to a galactocentric
%distance of 4\arcmin (22 kpc), the rotation amplitude is constant. 
 
The comparison with \citet{mcneil12} reveals that  the amplitude of rotation of PNe in NGC 1316 is significantly smaller (85$\pm$11)
than our rotation amplitude of the galaxy light in the same radial regime.  The PNe sample is plausibly biased towards
bright PNe, stemming from younger populations, while the galaxy light samples the entire, luminosity weighted range of populations present in NGC 1316.
Therefore this difference between PNe and galaxy light is suggestive of a kinematical difference between older and younger
populations, as the GC kinematics reflect it.

%\begin{table}[]
%\caption{Rotation amplitude for NGC 1316 in different radial bins. }
%\begin{center}
%\begin{tabular}{ccccc}
%\hline
%samples & total & $<$2.5\arcmin & 2.5\arcmin-3\arcmin& 3\arcmin-4.5\arcmin \\
%\hline
%A[km/s] &  115 $\pm$6 & 102$\pm$7 & 111$\pm$11 & 130$\pm$12\\
%\hline 
%\end{tabular}
%\end{center}
%\label{rotation}
%!TEX encoding = UTF-8 Unicode\end{table}
%

\begin{figure*}[t]
\begin{center}
\includegraphics[width=0.7\textwidth]{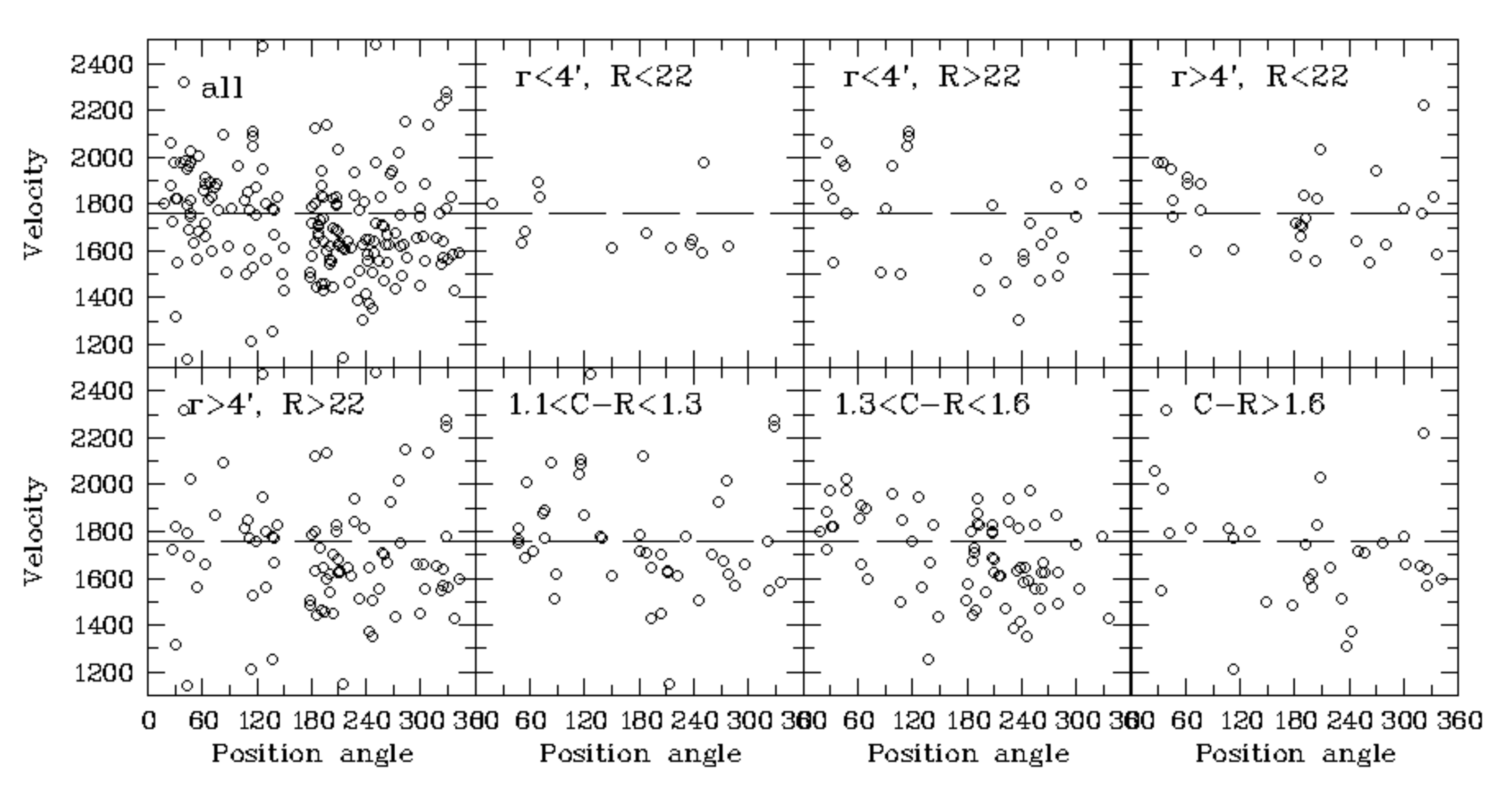}
\caption{Radial velocities vs. position angle for different subsamples. A clear rotation signal like that of the galaxy light is nowhere visible.
The sample with r$<$4\arcmin\ and R$>$22 mag, i.e. the faint bulge clusters, seems to show the clearest rotation signal.
 }
\label{fig:positionangle}
\end{center}
\end{figure*}

%\subsection{Remarks on rotation}

In  the projected light, the  inclination of a rotating structure is not the only uncertain point. In a mix of stellar populations with different kinematic properties,
some substructures may contribute to a rotation signal, others not. The rotation amplitude, as observed, may also be different at different wavelengths,
if the rotational behaviour depends on the population. 
%PNe? 

\subsection{Do the clusters rotate?}
The comparison of the galaxy velocities with the GC velocities is interesting. The rotation signal in Fig.\ref{fig:galrot} for the innermost positions is very
pure. Only for radii larger than 2\arcmin\ one finds velocities which do not fit into a strictly sinusoidal form. Part of the deviation might be caused
by low S/N, but the velocities of the bright GCs suggest that there are parts of the galaxy moving with velocities strongly deviant from the systemic
velocity or from a rotation pattern.
%some It is very clear that the GCs behave differently,
%even if low GC numbers prohibit  a direct comparison.   
%Most probably, there is a sub-population of GCs which corresponds exactly to the field population, but we will not identify this.

The rotation signal of the galaxy light is  fundamentally different from a rotation signal of GCs. The measured velocity of the
galaxy light is the luminosity-weighted mean along the line-of-sight without a-priori knowledge of the population which is responsible
for creating a rotation signal. The  
%average over  many stars, where the velocity dispersion cancels out.
 GCs, on the other hand,  play the role of single stars without the possibility to remove the velocity dispersion, so that one does
not expect such a clear rotation as measured in the galaxy. We also expect a contaminated rotation from the fact that velocities 
and locations can be related, e.g. in the southern L2-structure.  
Fig.\ref{fig:positionangle} shows the radial velocities vs. position angle for
the entire sample and some subsamples, selected according magnitude, radius, and colour.  The entire sample shows that
intrinsically we find many clusters at low  velocities  and positions angles larger than 180$^\circ$. This crowding is in part
due to the objects populating the velocity peak at 1600 km/s, which seems to be related to the L1-feature of \citet{schweizer80} 
(compare Fig.\ref{fig:1316sub} ).
 The clearest 
rotation is seen for the sample consisting of GCs fainter than 22 mag, and closer than 4\arcmin\ to the centre.
A fit to this sample reveals $a_0$ = 120$\pm$64 km/s and $\phi_0$=-1$\pm$64$^\circ$, demonstrating a large uncertainty.
At least it is consistent with the bulge rotation.
 All other GC samples
do not obviously rotate.

\section{Dynamical remarks}
\label{sec:dynamics}

Given the kinematic complexity, including the non-negligible rotational support, the mix of GC populations and uncertain three-dimensional structure, a proper dynamical analysis presently
is beyond our possibilities.  However, a few remarks are adequate. The first remark to be made is that the kinematical data for the
stellar population appear   not to be entirely consistent in the literature. \citet{donofrio95} measured a central velocity dispersion of 
260 km/s, in good agreement with \citet{bosma85}. The velocity dispersion then declines towards larger radii, reaching 150 km/s at
50\arcsec, while in Bosma et al.'s work, this decline is shallower and  reaches 150 km/s  at 80\arcsec. \citet{arnaboldi98} give a
central dispersion of only 200 km/s with a decline to 140 km/s at 60\arcsec. There are also asymmetries with respect to the center, particularly
pronounced along the minor axis.  On the other hand, \citet{bedregal06} find as well  a high central dispersion of about 260 km/s, but the
decline is much shallower and consistent with a constant velocity dispersion of 200 km/s between 50\arcsec(4.3 kpc) and 150\arcsec (12.9 kpc) along the major axis. Since the
VLT-data used by Bedregal et al. apparently have the highest S/N,  we adopt in the following their kinematics. There is more agreement regarding the LOS
velocities for which we adopt the Bedregal et al. values as well.
 
To represent the GC velocity dispersions, we avoid to use the full sample because of the bright cluster kinematics, which apparently do not fit to a Gaussian.
For a good sampling of the fainter clusters, we choose the limit R$>$21.5 mag to maintain a reasonable statistics and bin widths of 1.5\arcmin\ with an overlap
of 0.5\arcmin. These are the open circles in Fig.\ref{fig:models}.  Looking at Fig.\ref{fig:disp}, it is not surprising that deviations from a radially constant dispersion occur.
The sampling of velocities is far from being ideal.

\subsection{A spherical model}

In spite of all shortcomings, it is interesting to present   spherical models. Firstly, it can be compared to the model of \citet{mcneil12}, based on PNe. Secondly,
we can discuss the global characteristics of a dark halo without aiming at precision.  Thirdly, 
we can use our new photometric surface brightness profile in the R-band  from Paper I. Our model  uses the non-rotating spherical Jeans-equation, as do
 \citet{mcneil12}. The Jeans-formalism was presented in many contributions, we refer the reader to \citet{mamon05} and  \citet{schuberth10,schuberth12}. The surface brightness
 of NGC 1316 in the spherical approximation is well represented by a "beta-model", which can be deprojected analytically (e.g. \citealt{schuberth12}).
 %This spherical deprojection  
 %underestimates the real luminosity densities, unless the of an elongated configuration. The deprojected luminosity profile is 
 
 %$$   $$
 
 We then assign an M/L-ratio (R-band) and calculate the projected velocity dispersions, adding a dark halo to the baryonic mass. We use the formulas given by 
 \citet{mamon05}.  We choose a logarithmic halo (hereafter log-halo) with asymptotic circular velocity $v_0$ and core radius $r_0$, given by
\begin{equation}
 v_{\rm log}(r) = v_0 r/\sqrt{r_0^2+r^2}.
 \end{equation}
  
 %To compare the dark halo with predictions from cosmologically motivated simulations, we choose an NFW halo of the form  
% \begin{equation}
%M_{\rm{NFW}}(r)= 4 \pi \varrho_s r_s^3 \cdot \left( 
%\ln \left( 1 + \frac{r}{r_s} \right) -
%\frac{\frac{r}{r_s}}{1+ \frac{r}{r_s}}
%\right) \ ,
%\label{eq:massnfw}
%\end{equation}

\begin{figure}[h]
\begin{center}
\includegraphics[width=0.5\textwidth]{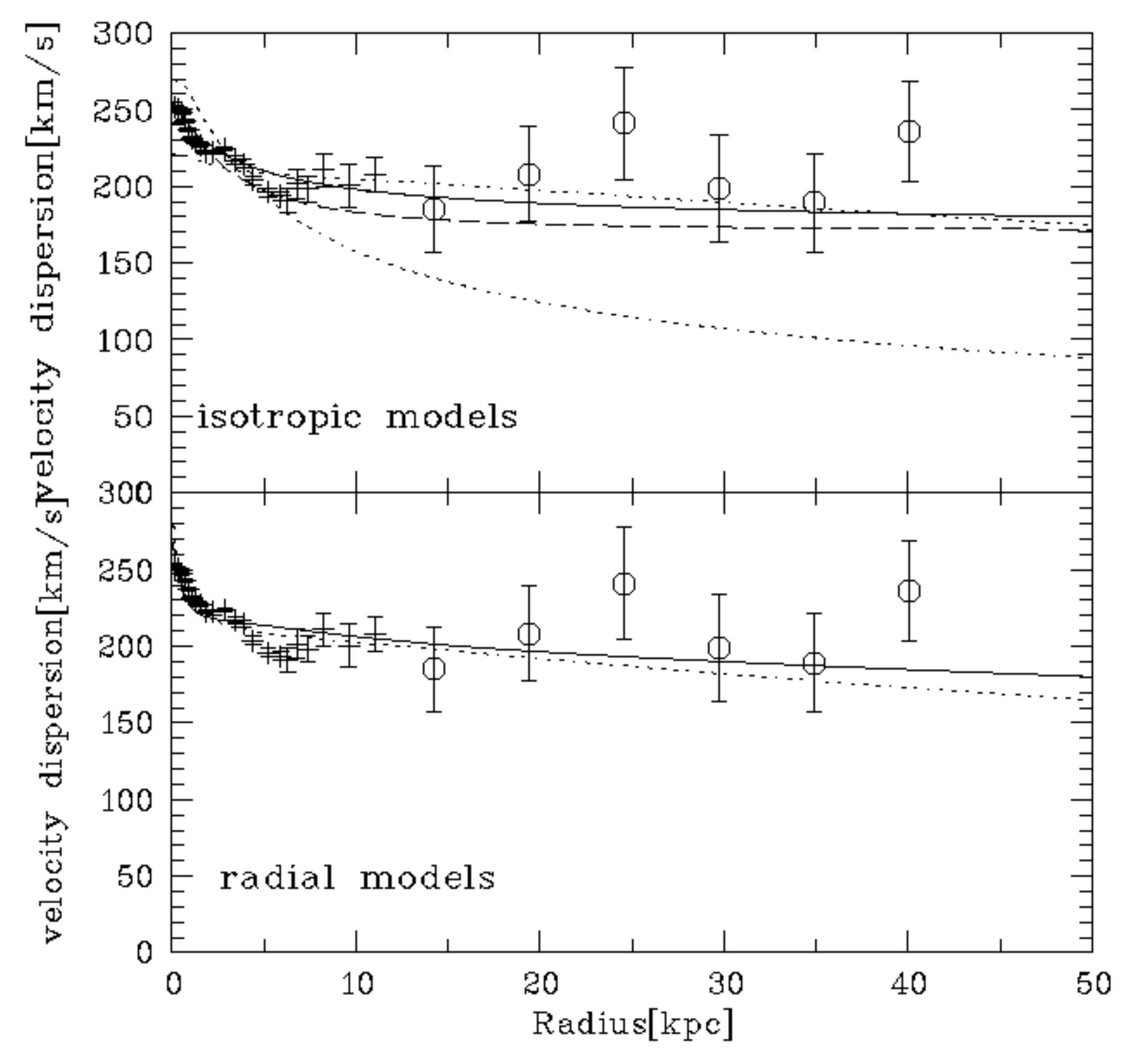}
\caption{In both panels, the crosses mark velocity dispersions from \citet{bedregal06}. The six open circles denote
velocity dispersions of clusters fainter than R=21.5 mag. Upper panel:  Some isotropic models.   The lower short-dashed line is a model
without   dark
matter and uses $M/L_R$=3.5. The solid line is a log-halo with  the parameters $v_0$=300 km/s and $r_0$=0.5 kpc and uses
$M/L_R$=1.3. The upper short-dashed line is a NFW-type halo with  $M/L_R$=2.0. See the text for its   parameters.  
The long-dashed line is a MONDian model with $M/L_R$=2.5.  Lower panel: Models with the radial anisotropy of \citet{hansen06}. The
solid line is a log-halo with the same parameters as above, but with $M/L_R$=2.0. The short dashed line is the same NFW halo as above to
show that $M/L_R$=2.0 results in a too high central velocity dispersion, but a stellar $M/L_R$ lower than 2.0 is difficult to justify.
 }
\label{fig:models}
\end{center}
\end{figure} 

%Since the data and all simplifications do not permit to strongly constrain the model parameters, the aim is not to find
%the best models, but to give an impression of the characteristics of the dark halo and also generally confirm what \citet{mcneal12} have been found.  

We first consider the simple case of  isotropic models, which are shown in Fig.\ref{fig:models} (upper panel). This maximizes the stellar M/L-value with respect to any radial anisotropy. To reproduce the central velocity dispersion of 250 km/s without dark
matter, one needs an
$M/L_R$-value of 3.2, distinctly higher than the value of 2.5, we advocated in Paper I, and of course much higher than the value of \citet{mcneil12} ($M/L_R\approx 1.7$), whose best fit model is radially anisotropic. This high value has no support by any dynamical study (see the discussion of \citealt{richtler11b}). Assuming solar metallicity and a
Chabrier-IMF, it would correspond 
to an age of 5.5 Gyr as a single stellar population \citep{marigo08}.
%But as Fig.\ref{fig:models} shows,
 %a dark halo  is necessary also in this case of a high M/L.
  
To minimize the dark halo, we want to keep the stellar $M/L$ as high as possible. On the other hand, a relatively small $r_0$ is needed to model the rapid decline
of the stellar velocity dispersion. One has to lower  $M/L_R$ until 1.3  to permit a log-halo with $r_0$=0.5 kpc and $v_0$=300 km/s.

% which is shown in Fig.\ref{fig:models}.
However, the central density of a log-halo is 
\begin{equation}
\rho_0 = 3 (v_0/r_0)^2/(4 \pi G).
\end{equation}
which means for the present halo a central density of  20 $M_\odot$/pc$^3$. The "surface density" \citep{donato09} is $\rho_0 \times r_0 \approx 10^4 M_\odot/pc^2$. These values
are not realistic in that they are much too high. Typical central densities of dark matter in massive elliptical galaxies are approximately $0.4 M_\odot$/pc$^3$
\citep{richtler11b}.
We come back to that in the discussion.

For comparison, we give a MONDian halo under isotropy with the MONDian circular velocity  

\begin{equation}
 v_M = \sqrt{v_N^2(r)/2 + \sqrt{v_N^4(r)/4 + v_N^2(r)  a_0 r}},
\end{equation}

where we adopt $a_0 = 1.35\times10^{-8} cm/sec^2$ \citep{famaey07}. Such halo needs $M/L_R$=2.5.
% and seems to underestimate a bit the
%dispersion values. 
%(whatever that means).

 However, merger simulations rather indicate modest radial anisotropies. We use the findings of \citet{hansen06} that the resulting anisotropy of stars in their merger simulations is related to the logarithmic slope of the three-dimensional stellar mass 
distribution by
\begin{equation}
   \beta = 1 -1.15 (1+slope(r)/6).
 \end{equation}  
     For our photometric model,  $\beta$ reaches a constant radial anisotropy of +0.4 at about 5 kpc.  A  good approximation for this relation
is the anisotropy profile considered by \citet{mamon05}: $\beta = 0.5(r/(r+r_a))$, $r_a$ being a scale radius with some low value.  Adopting this kind of anisotropy, we
can conveniently apply the formalism given by \citet{mamon05}. A consequence of the anisotropy is to lower the
stellar M/L-ratio to comply with the central velocity dispersion, which is boosted by the projected radial contributions. In the outer parts, the radial anisotropy results in
a lower projected velocity dispersion.   So we have to lower the M/L even more (which contradicts all existing dynamical and population evidence) or work with a more  
realistic value and reduce drastically the dark matter content in the inner region. 

The parameters $M/L_R$=2, $r_0$=5 kpc, and $v_0$=300 km/s do a good job. This halo is shown in Fig.\ref{fig:models}. Its central density is $0.2 M_\odot/pc^3$ and the
surface density is $10^3 M_\odot/pc^2$, which are consistent with values for massive elliptical galaxies. This is more or less a halo of the kind which \citet{mcneil12} derived from
planetary nebulae.

 To enable the comparison with the dark matter densities of elliptical and spirals  by \citet{napolitano10}, we also give dark matter profiles of the NFW-type:

\begin{equation}
\rho = \frac{\rho_s}{r/r_s (1+r/r_s)^2}
\end{equation}

with $\rho_s$ and $r_s$ being a characteristic density and a scale radius, respectively. With $r_s$ =17 kpc and $\rho_s = 0.028 M_\odot/pc^3$ in combination with $M/L_R=2.0$, one has a good
representation in the isotropic case (Fig.\ref{fig:models}, upper panel). For the anisotropic case (lower panel), we use the same parameters to show that the central velocity dispersion becomes  
too high and the M/L-value has to be lowered. The mean density within an effective radius of 68.9\arcsec (Paper I, appendix) or 6 kpc is 0.079 $M_\odot/pc^3$. Comparing this value with
Fig.9 of \citet{napolitano10} shows that it is a typical value for a massive elliptical. We comment on this  further in the discussion.

Looking at other halo shapes  is not worthwhile, given the observational constraints and model restrictions. They will differ in details,
 but not in the main conclusions which we reserve for the discussion (see \ref{sec:halos}).
   
\section{Discussion}

\subsection{The GC colour distribution within the globular cluster system}

It is very satisfactory that in the colour distribution of the ''pure'' GC sample,  the same features appear as in the photometric sample, namely a peak
at C-R$\approx$1.4 and a peak at  C-R$\approx$1.1.  This bimodal appearance of the colour distribution in NGC 1316 has  little to do with the bimodality, which is found in the GCSs of giant ellipticals, where the blue peak (C-R $\approx$1.3) consists of metal-poor, probably accreted clusters and the red peak of metal-rich
clusters (C-R $\approx 1.7$) formed with the majority of the metal-rich field population of the host galaxy (e.g. \citealt{richtler13}). Although ages have been spectroscopically determined
only for a few of the brightest clusters \citep{goudfrooij01b} of the red peak,  clusters as young as 0.5 Gyr can be identified by photometry alone,
provided that a radial velocity is available which excludes the nature as a background galaxy or as foreground star. The detections of GCs bluer than C-R=1.0 are serendipitous
and more objects remain to be discovered.  
The colour interval between the red peak and C-R$\approx$1.0
could be populated by metal-poor, old clusters, but since clusters around 0.5 Gyr definitely exist, we would also expect clusters  with ages between 
2 Gyr (the red peak) and 0.5 Gyr.  The colour variation among the  brightest clusters might indicate the duration of a period with a high star formation
rate, but high S/N spectra are necessary to investigate this in more detail, as well as to find out the fraction of old, metal-poor clusters. 
One notes  the absence of very bright clusters in the blue peak, but
 we designed our masks 
leaving out objects brighter than R=20 mag   and bluer than C-R $\approx$1.0 (because we did not expect such bright and blue objects). 
%Clusters brighter than R=20 mag in 
%The detections of GCs bluer than C-R=1.0 are therefore serendipitous
%and more objects remain to be discovered.
%  However, from the photometry we do not expect many. We do not have yet a direct proof, how many old, metal-poor clusters are among the GCS, but there should
%not be too many. Our spectra do not have the required S/N for a proper age determination, but since clusters as young as 0.4 Gyr are clearly identified,
%there should be a lot 
%In the cases of these objects one 
%We do not know of any other case, where 

\subsection{Comparison with planetary nebulae}
\label{sec:pne}
Although it is beyond our scope to discuss in detail the kinematics of PNe presented by \citet{mcneil12}, some remarks
on the comparison between  GCs and PNe are appropriate. The question is to what level are the kinematic properties of PNe and GCs
comparable? The youngest GC populations will have no PN counterparts and the main population of bright PNe will stem from
intermediate-age populations \citep{buzzoni06}. The velocity dispersions of the total samples of GCs and PNe agree within the uncertainties.

\begin{figure}[h]
\begin{center}
\includegraphics[width=0.5\textwidth]{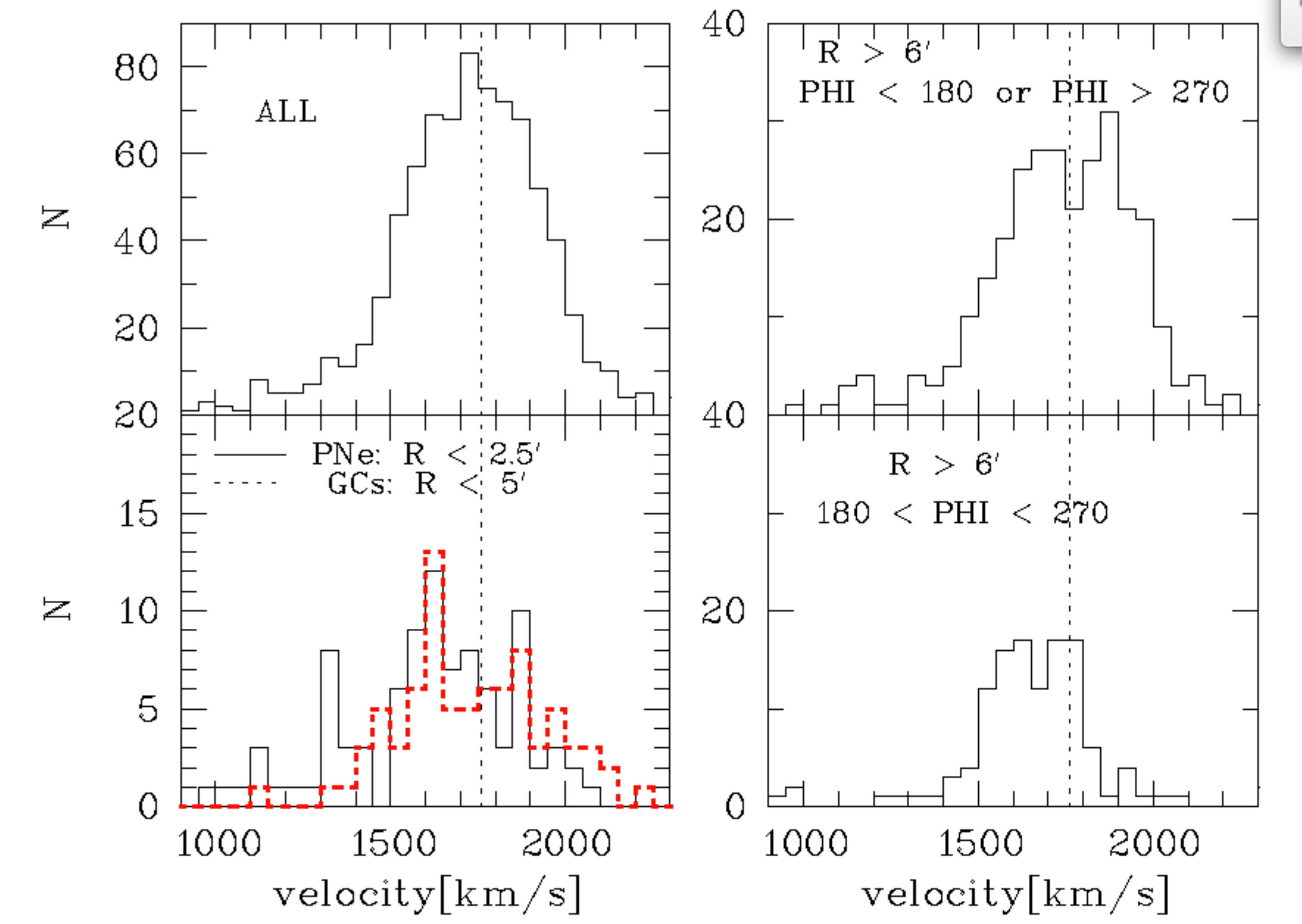}
\caption{For comparison with the GCs, some PNe samples are shown. Upper left panel: Entire sample for comparison. Lower left panel: inner 
PNe (solid line). Note the striking similarity with the globular clusters (dashed red line). Upper right panel:  outer PNe covering a position angle interval of 270$^0$ with the exception of the south-western quadrant. Here the velocity distribution is symmetric with respect to the systemic velocity. This histogram is meant as a supplement to the lower right panel.  Lower right panel: the south-western quadrant. Like in the case of globular clusters (Fig.\ref{fig:veldistri}), the velocity distribution is shifted
 towards lower velocities. The peak at 1600 km/s is not as pronounced as in the case of globular clusters.}
\label{fig:pne}
\end{center}
\end{figure}

Fig.\ref{fig:pne} shows velocity distributions of PNe, using the list published by \citet{mcneil12}. The upper left panel shows the total
sample, which is of course also presented by McNeil-Moylan et al., but here the binning is different and the peak near the systemic
velocity of NGC 1317 is not visible.  The PNe inside a radius of 2.5\arcmin\  (lower left panel) show  a picture similar to the bright inner GCs.
We cannot compare with the GC population within the same radius, but the comparison with GCs within 5\arcmin\  (which radius is needed
for producing comparable numbers), shows a perplexing agreement. The GCs are depicted by the dashed histogram. The peaks at 1600 km/s and 1900 km/s
are exactly reproduced. This agreement vanishes, when PNe at larger radii are included. 
 Because of the
galaxy's   brightness in the central parts, one may assume that the PNe are particularly bright and belong in their majority to the 2 Gyr population.
 % As do the GCs in Fig.\ref{fig:col_vel}, the PNe avoid the systemic velocity of NGC 1316 and are shifted to lower radial velocities.
  %The similarity to the velocity distribution of GCs in the colour interval 1.3$<$C-R$<$1.6 (which {\it bona fide} belong to a
  %2 Gyr population) is perplexing (Fig.\ref{fig:veldistri}). The dashed line is the histogram of GCs within 5\arcmin\ .  
  %However, we have hardly GCs in this inner region and 
  We interpreted the velocity peak
  of GCs at 1600 km/s as a signature of a disk-like distribution of clusters in the outer south-western part of NGC 1316, which is dominated
  by Schweizer's L1 structure. 
This view still holds, when looking at the lower right panel of Fig.\ref{fig:pne}, which is the outer south-western quadrant of the PNe distribution.
The shift of the distribution and the peak at 1600 km/s are clearly visible. The upper right panel for comparison shows the complementary distribution which is well centered
on the systemic velocity.  But then one would not expect that the inner peak at about the same velocity is due to the same feature, unless
there is a disk-like distribution of PNe over the entire galaxy.  This issue remains open for further investigation.

\begin{figure}[h]
\begin{center}
\includegraphics[width=0.4\textwidth]{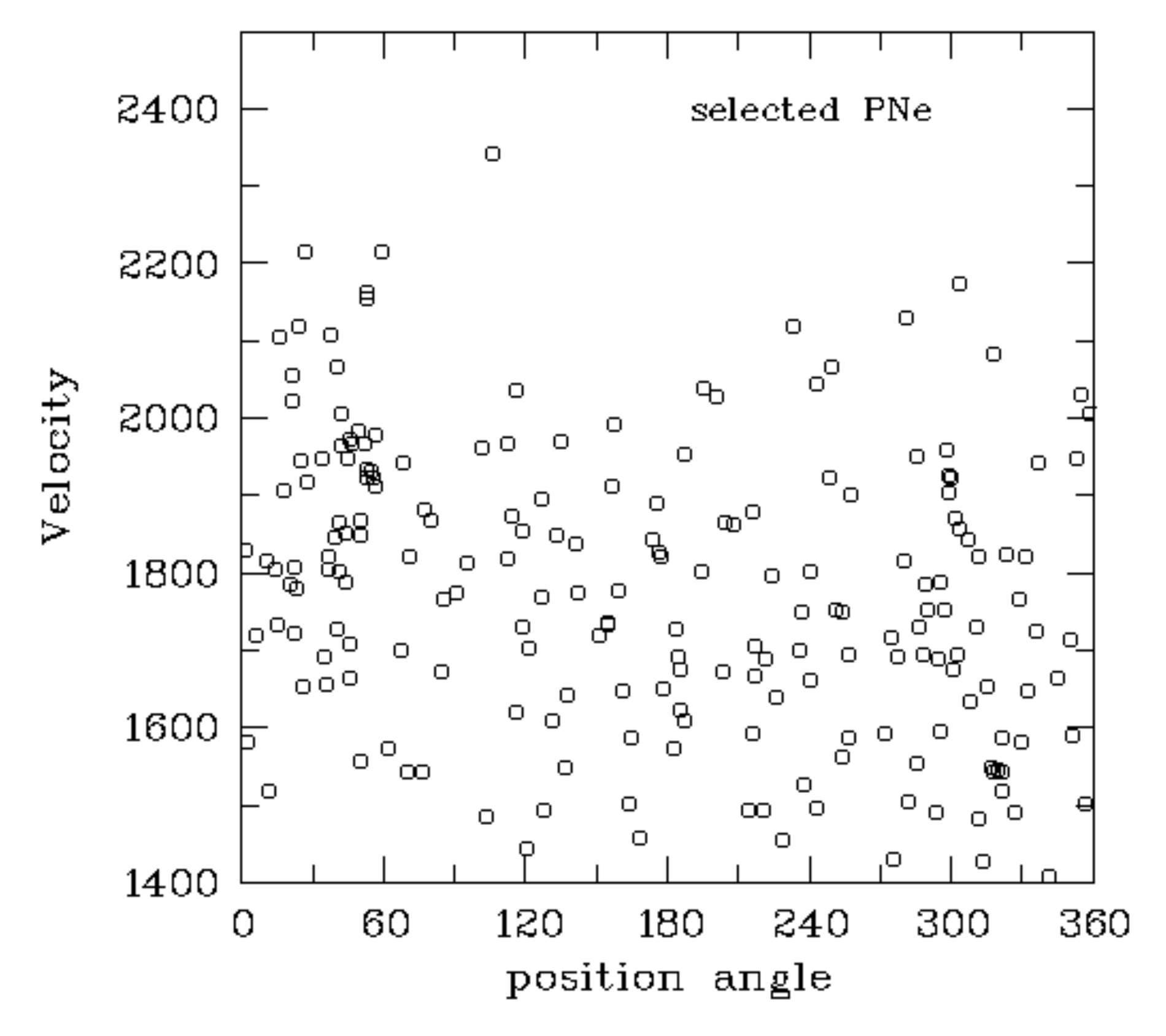}
\caption{Selection of the PNe from \citet{mcneil12}: distance between 2.9\arcmin\ and 4.3\arcmin. This plot shall demonstrate the similarity
with Fig.\ref{fig:galvel} in that there is  a striking population of high PNe velocities at  position angles less than 60$^\circ$.  The velocity distribution of PNe might represent
the velocity distribution along the line-of-sight, which is measured with luminosity weights by the integrated light.  }
\label{fig:pnegal}
\end{center}
\end{figure}

 A further interesting point  emerges from comparison of Fig.\ref{fig:galvel} with the distribution of PNe velocities. Fig.\ref{fig:galvel}
shows that the galaxy velocities reach quite high values for the largest distances and position angles in the range between 0$^\circ$ and 90$^\circ$. 
The same can be seen in the PNe velocity distribution. The highest PNe velocities are found in the radial distance range 2.9\arcmin $<$R$<$4.3\arcmin.
If we select PNe with these distances, we can plot Fig.\ref{fig:pnegal} which shows velocities vs. position angles. The PNe population between 0$^\circ$ and 90$^\circ$
is striking. It represents the stellar population in the lines-of-sight along which the galaxy velocities are measured. Since the galaxy velocities are luminosity-weighted mean
values,  their  nature  is difficult to analyse without the knowledge of the full velocity field.  But velocities as high as 2200 km/s are hardly rotation velocities and probably
related to the merger history.
%It would be very interesting to have the full velocity field. 

The velocity distribution of GCs in the bulge region is very similar
to the velocity distribution of PNe, but at larger radii the velocity
dispersions differ significantly. The velocity dispersion of GCs is
more or less constant, while that of the PNe starts to decline at a
radius of 200\arcsec\ (about 17.3 kpc). Since both trace the same mass, any difference
probably is  due to a difference in the three-dimensional distribution.
Again, any reasoning must remain
speculative at this point due to the uncertainty regarding this
distribution and the detailed population properties.
The parent population of PNe is of intermediate age, while GCs cover
a larger range of ages. Even if we cannot identify old clusters, one
would reasonably assume their presence. Their spatial distribution will be more
spherical than that of the PNe, which trace the outer stellar structures.
The  GC sample  at larger radii will be also contain a higher proportion of old
clusters compared to the bulge population. We therefore suspect that the PNe will
contain a higher fraction of objects belonging to a somewhat flattened parent
population, which in a radial average naturally  results in a lower line-of-sight dispersion.

\subsubsection{High velocity offsets}
In the GC sample and even more pronounced in the PNe sample, one finds a few objects with astonishingly large radial velocity offsets to NGC 1316,
in the case of PNe up to 1000 km/s. As \citet{mcneil12} suggest, some of the extreme cases may be Lyman-$\alpha$ galaxies at z=3.
 
In the Milky Way, such velocities would be labelled "hypervelocities" and a  common interpretation is the acceleration by the supermassive black hole
(SMBH) in the galactic center (e.g. \citealt{brown12,hills88}). These objects are extremely rare. The SMBH in NGC 1316 is more massive and the
stellar density might be higher, but one would not expect to find these stars in appreciable numbers in a PNe sample (in this case, the kinematics
of PNe would perhaps not say much about the potential of NGC 1316).  Large velocity offsets have been found also in NGC 1399
 \citep{richtler04,schuberth10}, but NGC 1399 is in the center of the Fornax cluster, while there are no such large potential differences near NGC 1316.
 The other possibility is that radial velocities near 1000 km/s are recession velocities rather than peculiar velocities. The PNe would then belong to an
 intergalactic population. This has been suspected before in the case of GCs around NGC 1399 \citep{richtler03a}, but until now can be neither confirmed
 nor discarded.

\subsection{The dark halo of NGC 1316: Spiral galaxy or elliptical galaxy?}
\label{sec:halos}
As discussed in Paper I, all evidences point toward
% a massive spiral galaxy plus one or more less massive galaxies or two 
spiral galaxies as pre-merger components. In brief, the
arguments are: the old globular cluster system, although it cannot be  identified cluster-by-cluster, must be quite poor, not fitting to a giant elliptical galaxy. Moreover,
a simple population synthesis requires an intermediate-age population for the pre-mergers in order to reproduce the galaxy colour, not an old population. On the other hand, the dark halo  {\it does not show the characteristics of dark halos of spiral galaxies, but fits to massive elliptical galaxies.}

As first shown by \citet{gerhard01}, 
% (see also \citealt{tortora09}),
 the dark halos of spiral galaxies, when represented by logarithmic halos, have central densities significantly lower than those of elliptical
galaxies of comparable mass.  The estimated factors vary between 10 and 30.  These factors appear lower in the more recent work of \citet{napolitano10}. 
However, the central density of our low-density log-halo ($0.2 M_\odot/pc^3$) is, what we would expect
for an elliptical galaxy with a  stellar mass of about $2.5\times 10^{11} M_\odot$. The dark halo of \citet{mcneil12} has a central density of 0.12 $M_\odot/pc^3$, but
their model has a constant radial anisotropy of $\beta$=+0.4 and demands $M/L_B = 2.8$ which corresponds to about $M/L_R$= 1.7, a quite low value,
which would correspond to a single stellar population of 1 Gyr \citep{marigo08}.
As the authors say, these parameters may change with a more sophisticated modelling.  

However, one would expect that the collisionless merging of two dark matter halos
would {\it lower} the characteristic densities and not enhance them. 
%Speculation: a subcluster halo.

It is perhaps too early to call that a serious problem in view of the simplicity of the present approach. At least it points toward the possible existence of an
 inconsistency 
%strong contradiction
 in the context of $\Lambda$CDM  and 
adds to the many other  unsatisfactory findings, see the compilations of \citet{kroupa12} and \citet{famaey12}. 

That the MONDian interpretation with $M/L_R$=2.5 works quite well, is at least remarkable.   Again, it is  not necessarily a  strong point for MOND, given the equilibrium  assumption of an apparently
quite chaotic system. However,  NGC 1316 would not be the first  galaxy outside the world of disk galaxies, where MOND works well only with an M/L, which fits to their evolutionary
history. \citet{milgrom12} recently showed, that the isolated ellipticals NGC 720 and 1521 are consistent with the MONDian phenomonology. Both are late merger
remnants (as it is the  case with many isolated "ellipticals"; \citealt{tal09,lane13}) with lower M/L-values than for old, metal-rich elliptical galaxies. So it not only seems that
the MONDian phenomenology extends to elliptical galaxies (or in other words, that elliptical galaxies fall onto the same baryonic Tully-Fisher relation as
disk galaxies), but that from the LCDM point of view, the dark matter content depends somehow on the M/L-values of the stellar population, which would be intriguing.

\section{Summary and conclusions}
%After a photometric investigation of the globular cluster system of NGC 1316 (Fornax A) 
We present radial velocities of 177 globular clusters in NGC 1316 (Fornax A), obtained with mask spectroscopy using FORS2/MXU at the VLT. To these data, we
add  20 radial velocities from \citet{goudfrooij01b}. Moreover, we determined radial velocities for the galaxy background light at 68 locations
out to a radius of 4\arcmin.
We discuss the kinematical structure of the globular cluster system and use existing data in combination with the globular clusters
to present a spherical dynamical model, using the photometric model from Paper I.
 
The most important  findings are:
\begin{itemize}
\item The colour distribution of confirmed GCs is bimodal, showing the same peaks as in the larger, but contaminated photometric sample of GCs. To these peaks, we assign two
         epochs of particularly high star formation rate, one at 2 Gyr, one at 0.8 Gyr. Moreover, we confirm that there  are a few clusters as young as 0.4 Gyr. 
         
\item Globular clusters brighter than about $M_R \approx -10$ mag   avoid the systemic velocity, particularly so the bright clusters of the Goudfrooij et al. sample,
which are constrained to the inner 5 kpc.
%  In their majority they do
%kinematically not belong to the bulge population.
 Their field stellar counterpart might be an extended stream. In this case, one would expect many
more bright clusters at large distances.  The inner planetary nebulae show a stunningly similar pattern.

\item A striking peak in the velocity distribution at 1600 km/s is mainly populated by clusters outside the bulge in the south-western region
of NGC 1316. This peak may indicate  a disk-like distribution of star clusters. We suggest that they belong to the structure L1, which {\it bona fide}
is the remnant of an infalling dwarf galaxy. 

\item The velocity dispersion of GCs shows a clear dependence of brightness by getting higher for fainter clusters, reaching a value
of about 200 km/s.

\item Out to 3\arcmin, the galaxy light shows a clear rotation signal with a more or less constant amplitude of about 130 km/s.  We do not find any subpopulation
of GCs with a similarly clear rotation. At larger radii the velocities scatter significantly, but we cannot distinguish between a chaotic velocity field and large
errors due to low S/N-spectra.

\item Disregarding the question, whether spherical models are good approximations or not, we present logarithmic halos as the dark matter halos, which can reproduce quite well the kinematics of the stellar light and the globular clusters.
Valid parameters are $r_0 = 5 kpc$, $v_0$ = 320 km/s, corresponding to a central dark matter density of around 0.2 $M_\odot/pc^3$. This halo is quite
similar to the dark halo shown by \cite{mcneil12}. 
\end{itemize} 

Given the entire kinematic evidence, the GCS cannot be described by simple morphological parameters like it is the case with many giant ellipticals. The large
variety of ages (metallicities are  unknown), the uncertain three-dimensional arrangement, and the perhaps complex velocity field reflect the complex history
of kinematics and dynamics. 
%However, this is a first attempt and a complete velocity field would bear 

%This refers particularly to the dark halo. That 

%NGC 1316 could be a key object for  LCDM .

The present dark halo
%although its definition is still deficient and needs more work, 
shows high dark matter densities typical for a massive elliptical galaxy, although
all indications rather point to spirals as merger progenitors. After merging activity, one would expect the pre-merger dark matter densities
to be even lower.  This conflict  is perhaps resolvable with MOND. Whatever the truth, NGC 1316 and its dark matter halo is probably a 
key object in  the discussion of LCDM and alternate gravity theories. 
%Given the complex dynamical situation, it will be a difficult task, but with the future
%knowledge of the complete velocity field, it could be possible.

%In a distant future, NGC 1316 will be a red, early-type galaxy with a poor cluster system and if it has the appropriate trajectory pointing away from
%the Fornax cluster, an isolated elliptical galaxy.
\bibliographystyle{aa}
\bibliography{N1316.bib}
\begin{acknowledgements}
We thank the anonymous referee for valuable and helpful comments.
TR acknowledges financial support   from FONDECYT project Nr. 1100620, and
from the BASAL Centro de Astrof\'{\i}sica y Tecnologias
Afines (CATA) PFB-06/2007. TR also thanks ESO/Garching, where the revised version
was completed.
MG thanks UNAB/DGID for financial support. 
LPB acknowledges support by grants from Consejo
Nacional de Investigaciones Cient\'ificas y T\'ecnicas and Universidad
Nacional de La Plata (Argentina).
\end{acknowledgements}

\appendix
\section{Morphological remnants of dwarf galaxies}
\begin{figure}[h]
\begin{center}
\includegraphics[width=0.5\textwidth]{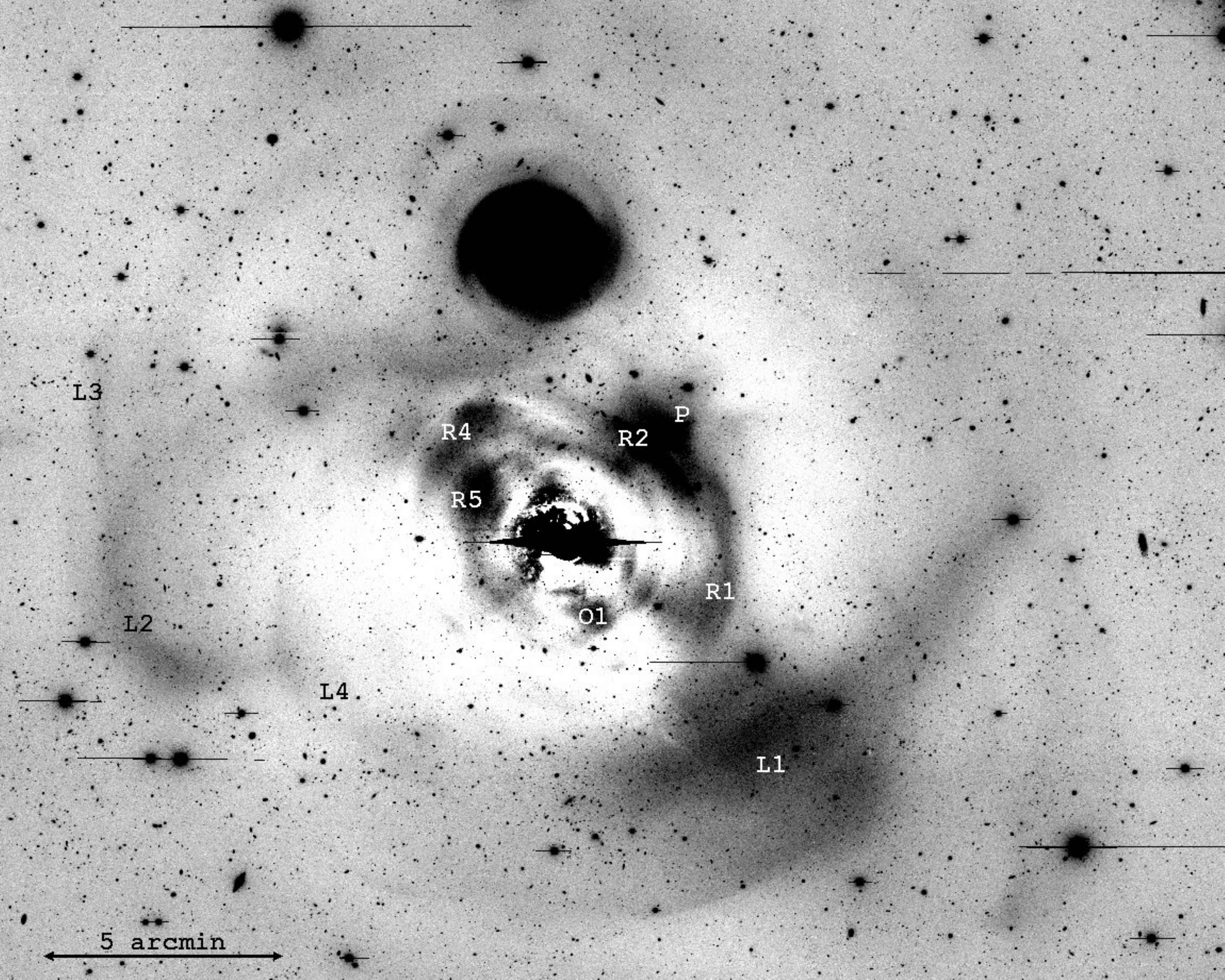}
\caption{Residuals in R after subtraction of a smooth elliptical model. North is up, east to the left. The designations are from \citet{schweizer80}. This global view shows Schweizer's L2 structure as a tidal tail. }
\label{fig:1316sub}
\end{center}
\end{figure}

We supplement our morphological remarks of Paper I by calling attention to structures which have either not been noted or mentioned with an unclear
interpretation.  In the following,
we use the designations of \citet{schweizer80} (see Fig.\ref{fig:1316sub}:L=loop, R=ripple, P=plume). Additionally, we introduce O1 as "object 1".

\begin{figure}[h]
\begin{center}
\includegraphics[width=0.5\textwidth]{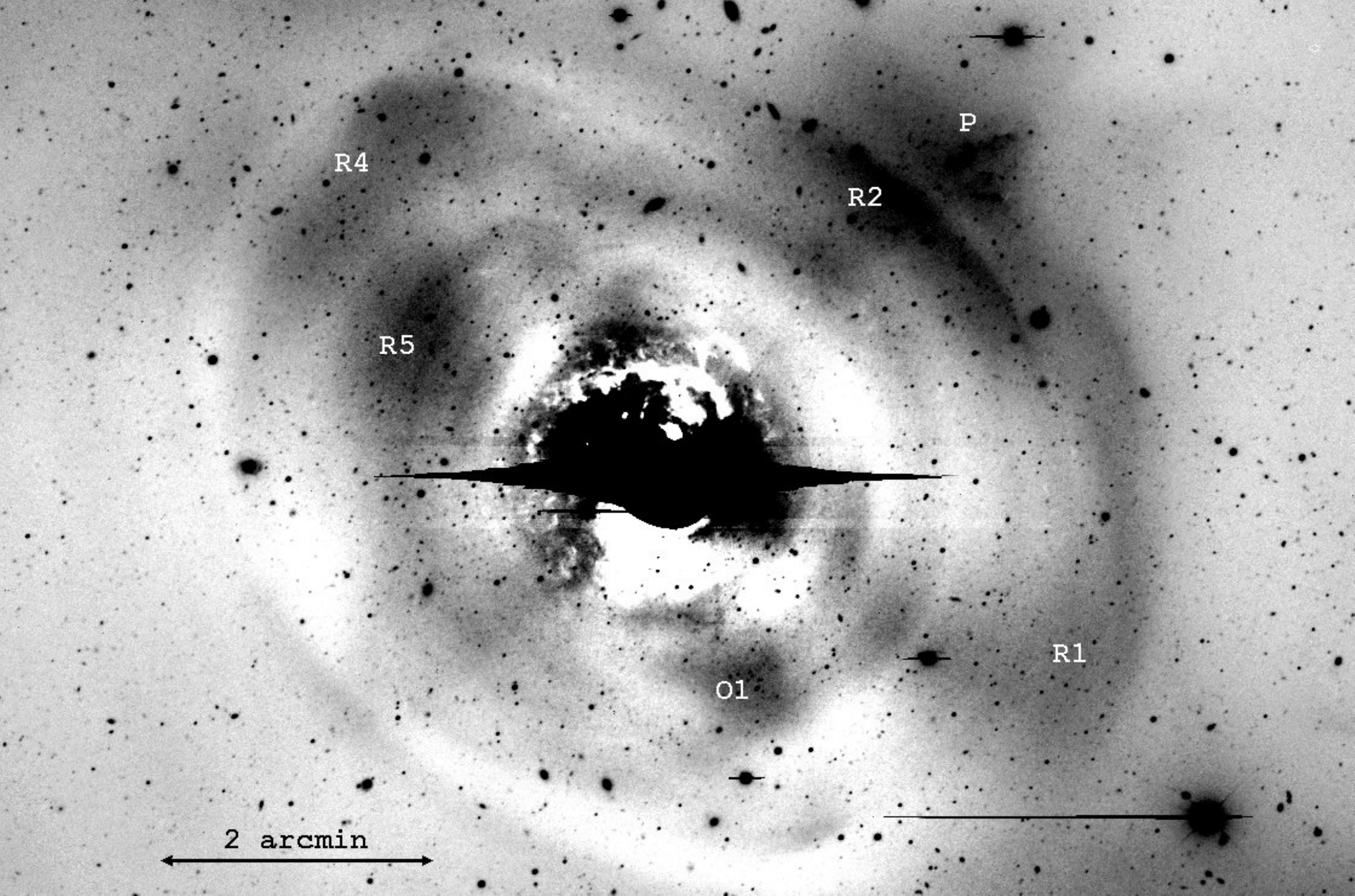}
\caption{Residuals in R of the inner part. The dynamical range is chosen to make the shells better visible. Note particularly the sharp
boundary of the shell R2. }
\label{fig:1316subinner}
\end{center}
\end{figure}

Already \citet{mackie98} showed 
residuals from an elliptical model, based on a photographic B-plate. Here we  see in more detail the  complex structures which become visible after the
subtraction of the elliptical model from Paper I. Fig.\ref{fig:1316sub} shows the wide field, while Fig.\ref{fig:1316subinner} demonstrates the structure in the 
inner parts. The shell system has been first described by \citet{schweizer80}. He identified 2 ripples on the south-western part, we see 
at least four.
Striking is the L2-structure in its full  extension. In the epoch of Schweizer's paper, computer simulations of galaxy interactions 
were just at their very beginning. Today we identify L2 as the long tidal tail of an infalling dwarf galaxy. Morphologically, it might be connected either
to NGC 1317 or to  Schweizer's ripple R2, which Schweizer suggested, but  if NGC 1317 would be related to this tidal structure, 
we would not expect such a seemingly undisturbed  spiral structure. 
%would not be so undisturbed as they appear. 
%do not show any signs of disturbation, particularly not the outer southern arm that should be most exposed to tidal influence?
  The  physical link to R2 is as well doubtful, given the quite different widths of the structures in the area of overlap. 
  % then one is tempted to  conclude that the dwarf galaxy traced by the tidal tail also is responsible 
 %for  the ripple whose trajectory can be followed over more than one full circle.   

The common wisdom today is that shells appear in simulations as caustics  in phase space after the infall of dwarf galaxies on a radial orbit into the  potential
of a larger galaxy (e.g. \citealt{sanderson13}). A morphological characteristic of these caustics  are the sharp outer boundaries which are the turn-around-points of stellar 
orbits.   Indeed, we find these sharp boundaries in NGC 1316 at the the well-known southern L1-feature, which accordingly has to be interpreted as the
remnant of a smaller galaxy. 
 But we find them also at some locations in the shell system, most
strikingly in the region of the "plume". It is difficult to see how {\it radial} orbits can play a role in this case. It might be of significance that the plume itself (which
probably is an infalling dwarf by its populations properties, see Paper I) has a radial structure.

The ripples are suprisingly  coherent. Following the outer shell clockwise, starting at the plume position, one is led on a spiral-like pattern until the inner region.
Interestingly, this path avoids  the ripple R2. 

Another pattern, which occurs in simulations (e.g. \citealt{sanderson13}),  are the T-like features. One conspicious example is located between L2 and
L4. 

\begin{figure}[h]
\begin{center}
\includegraphics[width=0.5\textwidth]{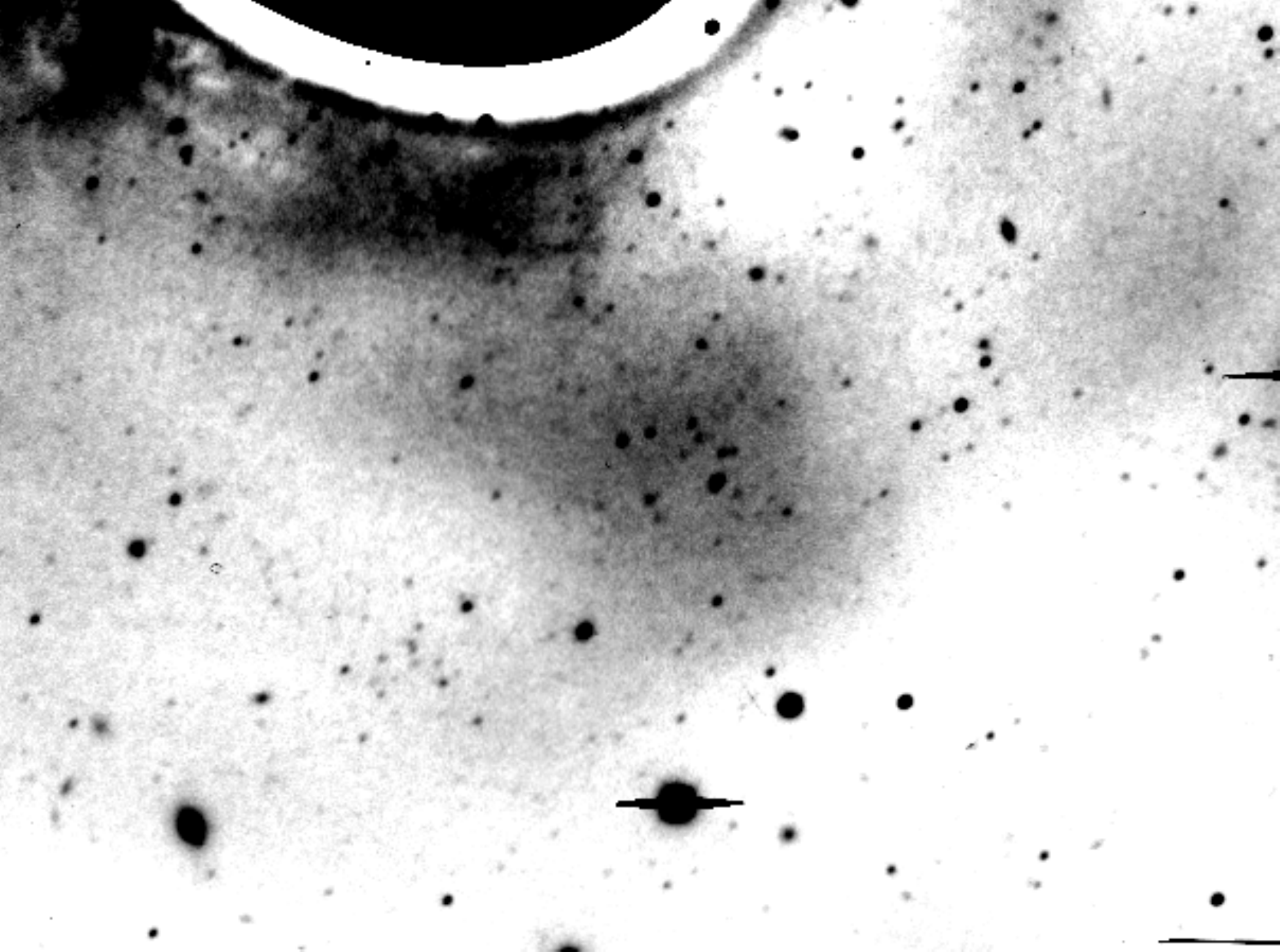}
\caption{The amplified region around O1. Note the apparently larger number of sources projected onto O1, which may be star clusters. One may speculate that
O1 is the remnant of a dwarf galaxy.  }
\label{fig:1316O1}
\end{center}
\end{figure}

Besides the "plume" region, the brightest residual, about 10\% of the underlying galaxy light is found at 50\arcsec west, 90\arcsec south, labelled ''O1''.
Fig.\ref{fig:1316O1} shows the region of this object. It is striking that O1 exhibits a larger density of sources than are found in its environment.

%\subsection{NGC 1317}
 
\subsection{NGC 1317}

\begin{figure}[h]
\begin{center}
\includegraphics[width=0.5\textwidth]{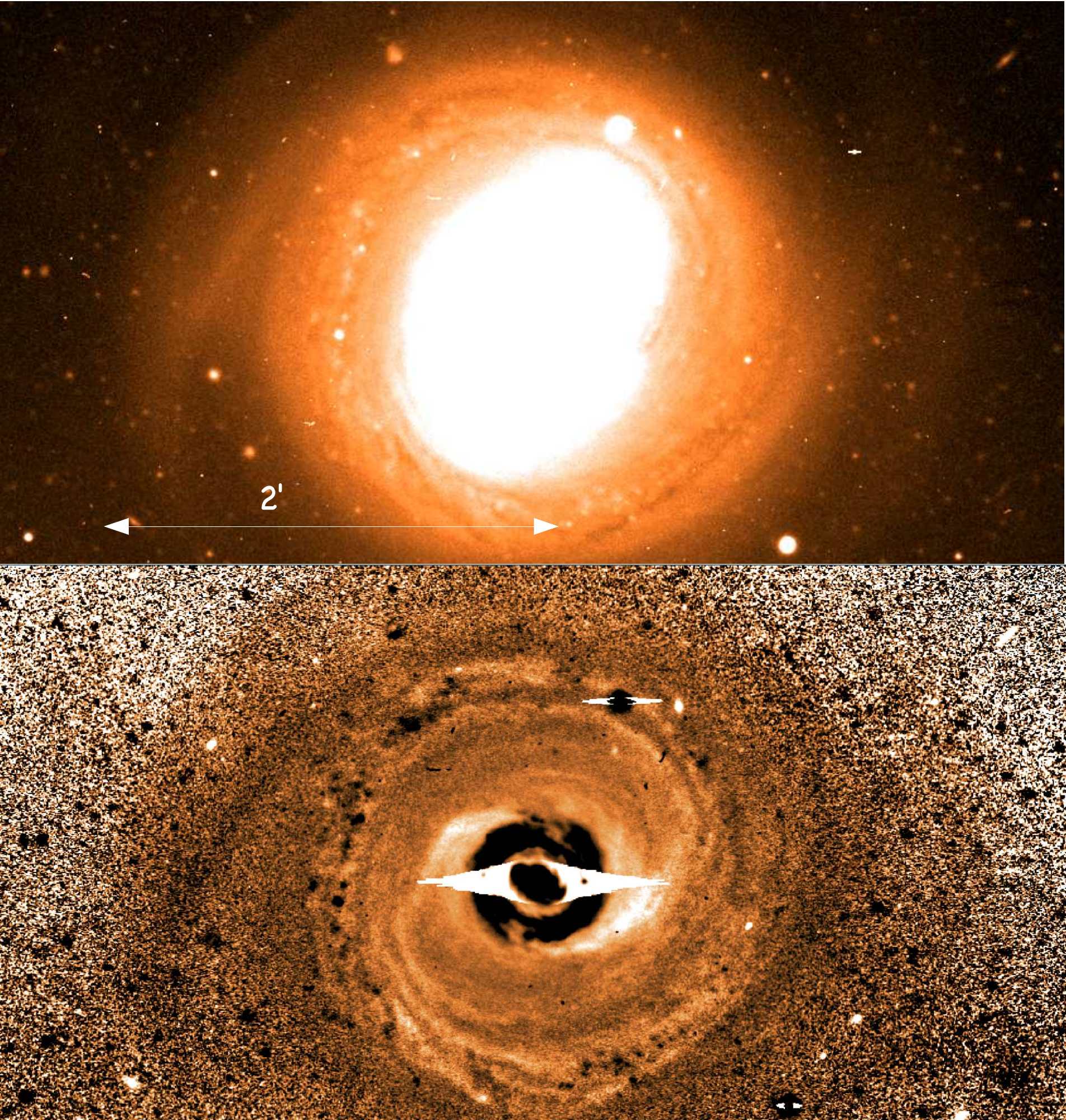}
\caption{The upper panel shows our C-image of NGC 1317, where due to the higher extinction the dust features are much better visible
than in the R-image.  North is up, east to the left.
The scale is valid for both panels. The lower panel is a colour map using  Washington C and Harris R (Paper I). The dynamical range of
the colour wedge is $ 1.3 < C-R < 2$. Blue is dark, red is bright.  Note the inner star forming ring, which appears black. The bright features
denote dust pattern. Outside the ring, more young populations are  visible as black spots arranged in a ring-like fashion on the eastern side.
 }
\label{fig:1317colour}
\end{center}
\end{figure}

The companion galaxy had to our knowledge never been the topic of a dedicated publication, although many highly interesting features can be identified. 
%NGC 1317 clearly is a peculiar galaxy.
Morphological  and stellar population aspects have been discussed by e.g. \citet{schweizer80,marcum01,papovich03} and particularly, using near-infrared filters, by \citet{laurikainen06}.
Here we briefly add a few morphological remarks on
NGC 1317 on the basis of a colour map, as has been done for NGC 1316 in Paper I, and HST-images, which to our
knowledge have not yet been shown in the literature.  NGC 1317 is a double-barred spiral galaxy with star formation occurring within a ring-like area.
In Fig.\ref{fig:1317colour}, the upper panel is a C-image, showing better the dust structures, while the lower panel is a C-R colour image (compare the colour image
of NGC 1316 in Paper I). It is striking that the ellipticity of these two images is so different.  We attribute this to the outer secondary bar \citep{laurikainen06}, which produces the ellipticity, but is not distinct in colour from the overall population.
 
 At higher  HST/WPC2 resolution, the "ring" is resolved into a  spiral structure with many  tightly wound arms,
 a point noted already by e.g. \citet{piner95} and \citet{lin08} for their simulations of star-forming rings in galaxies.
The inner radius is about 7\arcsec, the outer about 16\arcsec, corresponding to  604 pc  and 1380 pc, respectively, if we adopt for NGC 1317 
  the  distance of NGC 1316. This is also the region of H$_\alpha$-emission \citep{marcum01} and appears black (C-R $\approx$ 1.2) on our colour image. Outside this radius one cannot find coherent regions 
of star formation,  but sequences of blueish blobs/spots in the north-eastern  and south-western sectors. Because they trace the overall curvature, they 
very probably represent smaller scale HII-regions, indicating star formation on a lower level  than in the inner region. However, they are outside
the H$_\alpha$-map of \citet{marcum01}.  Interestingly, these two sectors
build part of a ring. 
%(commencing star formation at the outer Lindblad resonance?).
 Brighter colours in Fig.\ref{fig:1317colour} denote dust 
patterns.  It is intriguing that the dust, in the form of filaments, fills an area with a radius of about 5 kpc, but apparently without much  star formation, if any. In the very outer parts, NGC 1317 has
the appearance of a grand design spiral with two spiral arms dominating. These spiral arms, however, have colours comparable to the bulge
colour of NGC 1316, corresponding to populations with ages of about 2 Gyr. NGC 1317 might thus be a case for the longevity of spiral structures.  \citet{struck11} showed
how fast fly-by encounters can produce long-living density waves. Their simulations resemble quite well the appearance
of the outer structure of NGC 1317. Moreover, we point out the similarity with the multiple-ring galaxy  NGC 6782, which also presents
dust lanes resembling spiral arms. The dust is found between two rings of star formation.  In the model of \citet{lin08} for NGC 6782, the outer  ring appears between the corotation radius and the outer Lindblad resonance. In the case of NGC 1317, star formation in the outer ring might have died out and the
remaining HII-regions are only the debris of a previous prominent ring.  \citet{horellou01} note the unusually small HI-disk and mention the possibility that it has been affected by ram pressure, when transversing through the intergalactic hot medium. This then would have happened when the present intermediate-age population was young. Finally, we remark that if
%inner Lindblad resonance and the 4:1 strong bar potential is responsible for the peculiar
%appearance, which could be also the case for NGC 1317.
%to the galaxy and {\it but still a spiral stucture}. Apparently, the colour spiral structure is not determined by stellar population, but by the  The distribution of HI is but neutral hydrogen.    with the inner radius about  
 N1317 was in
% actual tidal interaction with 
exactly the same distance of N1316, one would expect strong tidal forces  which would not leave the disk intact and  the dust and probably molecular gas quiet with regard to star formation.
%\citet{horellou01} note the unusually small HI-disk and mention the possibility that it has been affected by ram pressure, when transversing through the intergalactic
%hot medium. This then would have happened when the present intermediate-age population was young, in line with the intermediate age population. 

%Inner star formation, but what populations at larger radii?
%The question is, how these structures could survive.
%Rather a outer spiral arm than something tidal
%Dust structure: newly formed?

\begin{figure}[h]
\begin{center}
\includegraphics[width=0.5\textwidth]{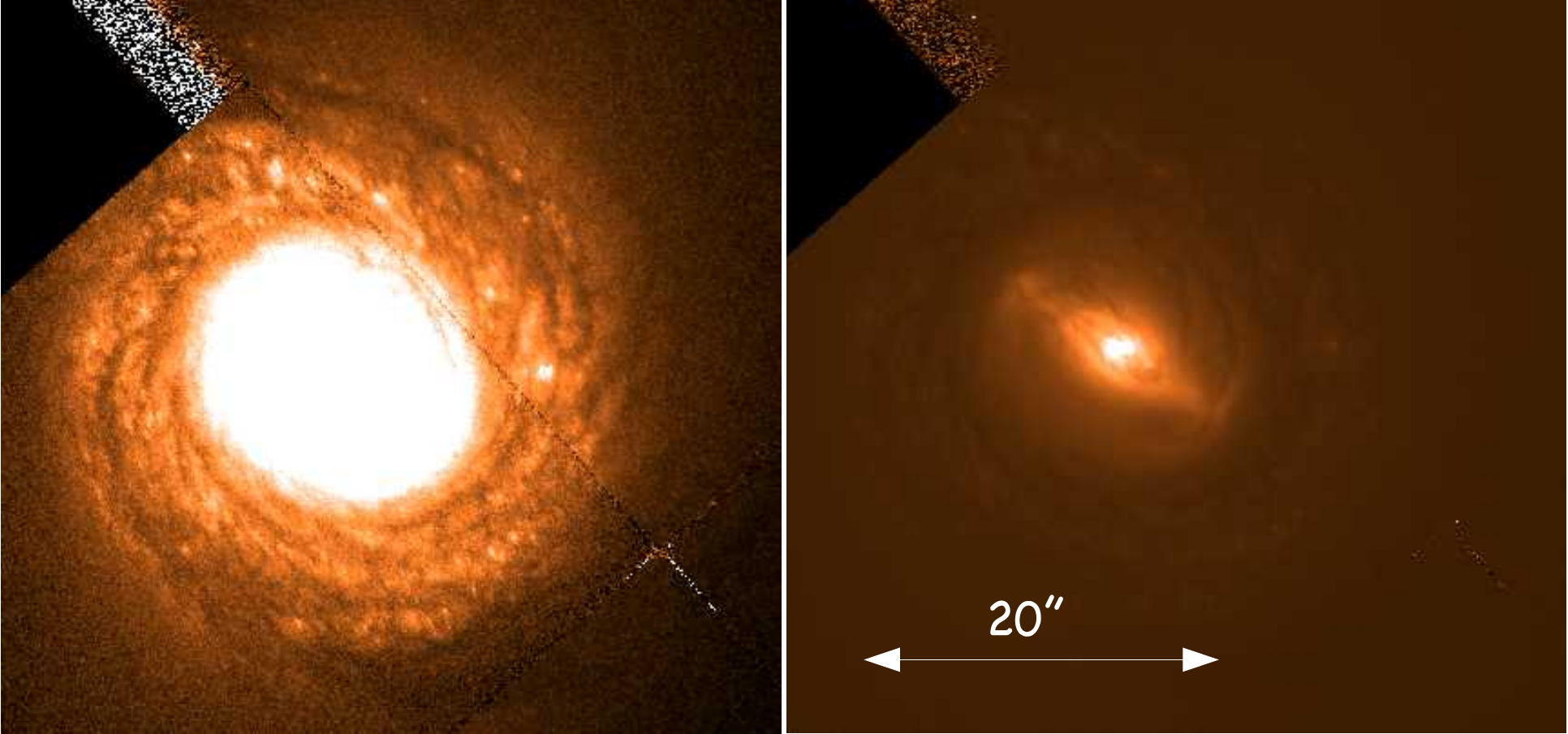}
\caption{ HST/WPC2-image of NGC 1317, F606W, 160s, exposure time. Program: 5446, PI: G. Illingworth. The left panel shows, how the
star-forming "ring" is resolved in many tightly wrapped up spiral arms. The right panel lets the inner bar become visible. North is up, east
to the left.
 }
\label{fig:1317hst}
\end{center}
\end{figure} 

\section{Tables}

Table \ref{tab:gcsample_points} lists all GCs which were selected as point sources in the photometry and thus have an entry in the photometric list
of Paper I \citep{richtler12e}. The columns are the catalogue number, the coordinates (J2000),  the R-magnitude, the colour C-R, the
heliocentric radial velocity and its uncertainty.  Table \ref{tab:gcsample} continues the GC list with those
objects, which are not in the point-source list of Paper I. Their catalogue number derives from the
internal catalogue in use and the photometric values of these objects appear only here.  The coordinates serve for identification purposes only.
Unknown magnitudes and/or colours appear as 99.99. The six double measurements (see Section \ref{sec:previous}) are included in Table \ref{tab:gcsample_points}. 

%\begin{table}
%\centering
%\resizebox{0.45\textwidth}{!}{
%\begin{landspace}
\begin{longtable}{ccccccc}
%\begin{longtable}{*7{p{0.05\textwidth}}}
\caption{Identification and radial velocities of globular clusters appearing in the catalog of Paper I.}\\
%\resizebox{0.25\textwidth}{!}{
\hline
\hline
Id  &  RA(J2000)  & Dec(J2000) & R & C-R &  rad.vel. &  error \\
\hline
%n1316\_gc00250 & 3:22:13.73 & $-$37:11:56.5 &  23.30 &  1.20 &  2020.0 &  61.0 \\
%n1316\_gc00280 & 3:22:23.21 & $-$37:08:26.0 &  22.05 &  5.07 &  1656.1 &  42.9 \\
%n1316\_gc00285 & 3:22:25.89 & $-$37:12:02.1 &  23.50 &  1.32 &  1869.2 &  56.7 \\
%n1316\_gc00286 & 3:22:26.02 & $-$37:05:17.1 &  23.30 &  1.33 &  1427.0 &  49.0 \\
n1316\_gc00250 & 3:22:13.73 & $-$37:11:56 &  23.30 &  1.20 &  2020 &  61 \\
n1316\_gc00280 & 3:22:23.21 & $-$37:08:26 &  22.05 &  99.99 &  1656 &  42  \\
n1316\_gc00285 & 3:22:25.89 & $-$37:12:02 &  23.50 &  1.32 &  1869 &  56 \\
n1316\_gc00286 & 3:22:26.02 & $-$37:05:17 &  23.30 &  1.33 &  1427 &  49  \\
n1316\_gc00293 & 3:22:27.50 & $-$37:17:54 &  23.34 &  1.44 &  1800 &  61  \\
n1316\_gc00315 & 3:22:33.72 & $-$37:18:07 &  22.87 &  1.63 &  1596 &  51  \\
n1316\_gc00319 & 3:22:35.58 & $-$37:18:09 &  21.89 &  1.61 &  1740 &  26 \\
n1316\_gc00328 & 3:22:39.10 & $-$37:19:55 &  22.99 &  1.25 &  2122 &  71 \\
n1316\_gc00341 & 3:22:42.73 & $-$37:21:33 &  22.41 &  1.41 &  1507 & 170 \\
n1316\_gc00383 & 3:22:52.85 & $-$37:10:26 &  22.68 &  1.12 &  1763 &  81 \\
n1316\_gc00384 & 3:22:53.00 & $-$37:12:48 &  22.94 &  1.31 &  1960 &  48 \\
n1316\_gc00410 & 3:22:58.85 & $-$37:08:49 &  22.86 &  1.61 &  1795 &  63 \\
n1316\_gc00420 & 3:23:01.40 & $-$37:08:34 &  22.74 &  1.00 &  1692 &  46  \\
n1316\_gc00461 & 3:23:12.47 & $-$37:19:04 &  22.85 &  1.26 &  1782 &  66  \\
n1316\_gc00478 & 3:23:16.69 & $-$37:20:23 &  22.57 &  1.46 &  1666 &  42 \\
n1316\_gc00485 & 3:23:18.96 & $-$37:15:04 &  23.03 &  1.39 &  1852 &  95 \\
n1316\_gc00500 & 3:23:22.69 & $-$37:14:57 &  22.54 &  1.83 &  1814 &  32 \\
n1316\_gc01061 & 3:22:01.53 & $-$37:15:12 &  22.59 &  0.97 &  2482 &  79 \\
n1316\_gc01075 & 3:22:02.63 & $-$37:13:20 &  23.22 &  1.31 &  1669 &  41 \\
n1316\_gc01108 & 3:22:06.91 & $-$37:12:34 &  21.89 &  0.96 &  1944 &  32 \\
n1316\_gc01117 & 3:22:07.70 & $-$37:15:50 &  22.30 &  1.37 &  1647 &  55 \\
n1316\_gc01177 & 3:22:15.44 & $-$37:06:07 &  21.68 &  1.16 &  1756 &  32  \\
n1316\_gc01178 & 3:22:15.87 & $-$37:02:18 &  21.83 &  0.37 &  1831 &  65   \\
n1316\_gc01193 & 3:22:17.90 & $-$37:16:12 &  22.96 &  1.82 &  1513 &  53 \\
n1316\_gc01214 & 3:22:20.35 & $-$37:06:22 &  22.17 &  1.85 &  1569 &  25   \\
n1316\_gc01214 & 3:22:20.35 & $-$37:06:22 &  22.17 &  1.85 &  1641 &  43  \\
n1316\_gc01230 & 3:22:22.00 & $-$37:09:49 &  22.78 &  1.32 &  1557 &  43 \\
n1316\_gc01273 & 3:22:26.89 & $-$37:11:59 &  22.76 &  1.57 &  1494 &  46 \\
n1316\_gc01278 & 3:22:27.23 & $-$37:18:00 &  23.25 &  1.47 &  1829 & 180 \\
n1316\_gc01282 & 3:22:27.52 & $-$37:12:23 &  22.35 &  1.29 &  1673 &  73 \\
n1316\_gc01304 & 3:22:29.58 & $-$37:13:26 &  21.49 &  1.56 &  1592 &  26 \\
n1316\_gc01315 & 3:22:30.70 & $-$37:17:22 &  23.32 &  1.18 &  1446 &  54 \\
n1316\_gc01324 & 3:22:31.54 & $-$37:15:26 &  21.25 &  1.46 &  1610 &  21 \\
n1316\_gc01350 & 3:22:34.75 & $-$37:17:41 &  22.94 &  1.09 &  1456 &  59 \\
n1316\_gc01389 & 3:22:39.07 & $-$37:20:15 &  22.59 &  1.56 &  1802 &  38 \\
n1316\_gc01462 & 3:22:46.10 & $-$37:09:44 &  21.67 &  1.45 &  1799 &  38 \\
n1316\_gc01524 & 3:22:50.48 & $-$37:13:02 &  22.70 &  1.49 &  1500 &  58 \\
n1316\_gc01530 & 3:22:51.15 & $-$37:08:43 &  23.03 &  1.45 &  1722 &  48  \\
n1316\_gc01540 & 3:22:52.36 & $-$37:09:13 &  22.91 &  1.68 &  1548 & 103 \\
n1316\_gc01586 & 3:22:56.56 & $-$37:13:54 &  22.52 &  1.16 &  2111 &  59 \\
n1316\_gc01623 & 3:22:59.72 & $-$37:11:26 &  20.91 &  1.18 &  1874 &  31 \\
n1316\_gc01652 & 3:23:03.59 & $-$37:11:53 &  23.21 &  1.10 &  2097 &  81 \\
n1316\_gc01695 & 3:23:09.21 & $-$37:15:32 &  23.20 &  1.47 &  1755 &  59 \\
n1316\_gc01748 & 3:23:16.64 & $-$37:21:29 &  22.61 &  1.58 &  1828 &  39 \\
n1316\_gc02665 & 3:22:01.23 & $-$37:16:01 &  23.16 &  1.14 &  1505 &  44 \\
n1316\_gc02736 & 3:22:06.74 & $-$37:10:58 &  23.63 &  0.94 &  2151 &  65 \\
n1316\_gc02755 & 3:22:08.05 & $-$37:11:37 &  22.02 &  1.69 &  1751 &  62 \\
n1316\_gc02786 & 3:22:11.12 & $-$37:09:05 &  22.31 &  1.03 &  1450 &  88 \\
n1316\_gc02802 & 3:22:12.58 & $-$37:13:27 &  22.51 &  1.19 &  1702 &  63 \\
n1316\_gc02807 & 3:22:13.23 & $-$37:04:47 &  23.13 &  1.17 &  1544 &  45 \\
n1316\_gc02812 & 3:22:13.46 & $-$37:12:12 &  22.88 &  1.09 &  1436 &  64 \\
n1316\_gc02834 & 3:22:15.18 & $-$37:03:10 &  23.07 &  1.04 &  1560 &  37 \\
n1316\_gc02840 & 3:22:15.51 & $-$37:13:09 &  21.81 &  1.56 &  1552 &  56 \\
n1316\_gc02851 & 3:22:17.01 & $-$37:15:37 &  22.32 &  1.36 &  1811 &  31 \\
n1316\_gc02855 & 3:22:17.27 & $-$37:14:59 &  22.43 &  1.61 &  1371 &  61 \\
n1316\_gc02874 & 3:22:19.18 & $-$37:14:18 &  21.77 &  1.05 &  1640 &  59 \\
n1316\_gc02877 & 3:22:19.49 & $-$37:02:37 &  21.52 &  1.22 &  1585 &  39\\
n1316\_gc02888 & 3:22:20.31 & $-$37:17:04 &  22.59 &  1.28 &  1611 &  47 \\
n1316\_gc02890 & 3:22:20.41 & $-$37:10:29 &  22.55 &  1.26 &  1656 &  39 \\
n1316\_gc02891 & 3:22:20.46 & $-$37:05:37 &  22.17 &  1.23 &  2252 &  70  \\
n1316\_gc02891 & 3:22:20.46 & $-$37:05:37 &  22.17 &  1.23 &  2279 &  34 \\
n1316\_gc02910 & 3:22:21.77 & $-$37:11:43 &  21.66 &  1.44 &  1626 &  51 \\
n1316\_gc02950 & 3:22:25.55 & $-$37:14:37 &  22.70 &  1.65 &  1307 &  37 \\
n1316\_gc02952 & 3:22:25.77 & $-$37:21:07 &  22.46 &  1.31 &  1542 &  63 \\
n1316\_gc02954 & 3:22:25.95 & $-$37:19:48 &  23.56 &  1.30 &  1698 &  42 \\
n1316\_gc02958 & 3:22:26.22 & $-$37:16:16 &  23.12 &  1.86 &  1644 &  57 \\
n1316\_gc02962 & 3:22:26.53 & $-$37:14:04 &  22.38 &  1.06 &  1557 &  61 \\
n1316\_gc02963 & 3:22:26.67 & $-$37:10:21 &  22.26 &  1.09 &  1883 &  61 \\
n1316\_gc02969 & 3:22:27.07 & $-$37:19:27 &  21.89 &  1.02 &  1553 &  59 \\
n1316\_gc02977 & 3:22:28.21 & $-$37:17:14 &  22.22 &  1.44 &  1626 &  43 \\
n1316\_gc02977 & 3:22:28.21 & $-$37:17:14 &  22.22 &  1.44 &  1683 &  41 \\
n1316\_gc02991 & 3:22:29.21 & $-$37:08:30 &  22.74 &  1.45 &  1778 &  39\\
n1316\_gc02997 & 3:22:29.42 & $-$37:13:22 &  21.14 &  1.45 &  1976 &  33 \\
n1316\_gc03002 & 3:22:29.84 & $-$37:11:51 &  22.93 &  1.29 &  1570 &  56 \\
n1316\_gc03004 & 3:22:29.87 & $-$37:13:53 &  21.58 &  1.53 &  1644 &  29 \\
n1316\_gc03019 & 3:22:31.10 & $-$37:13:02 &  20.07 &  1.49 &  1830 &  11 \\
n1316\_gc03025 & 3:22:31.46 & $-$37:18:17 &  23.58 &  1.85 &  1617 & 165 \\
n1316\_gc03027 & 3:22:31.61 & $-$37:16:07 &  21.57 &  1.72 &  2034 &  25 \\
n1316\_gc03030 & 3:22:31.70 & $-$37:12:37 &  20.83 &  1.55 &  1623 &  21 \\
n1316\_gc03033 & 3:22:31.72 & $-$37:13:41 &  20.83 &  1.49 &  1417 &  15 \\
n1316\_gc03041 & 3:22:32.12 & $-$37:13:10 &  22.40 &  1.79 &  1714 &  23 \\
n1316\_gc03044 & 3:22:32.39 & $-$37:13:57 &  21.96 &  1.26 &  1777 &  77 \\
n1316\_gc03046 & 3:22:32.59 & $-$37:12:48 &  22.14 &  1.44 &  1470 &  37 \\
n1316\_gc03055 & 3:22:32.95 & $-$37:12:15 &  21.28 &  1.17 &  1620 &  49 \\
n1316\_gc03067 & 3:22:33.69 & $-$37:15:35 &  23.20 &  1.53 &  1792 &  41 \\
n1316\_gc03090 & 3:22:35.24 & $-$37:16:02 &  22.12 &  1.63 &  1561 &  32 \\
n1316\_gc03096 & 3:22:35.35 & $-$37:17:58 &  22.50 &  1.19 &  1643 &  73 \\
n1316\_gc03103 & 3:22:35.83 & $-$37:20:08 &  22.49 &  1.42 &  1731 &  46 \\
n1316\_gc03122 & 3:22:37.32 & $-$37:16:05 &  21.88 &  1.40 &  1830 &  41 \\
n1316\_gc03132 & 3:22:38.12 & $-$37:15:36 &  22.19 &  1.11 &  1430 &  54 \\
n1316\_gc03133 & 3:22:38.13 & $-$37:19:31 &  22.68 &  1.40 &  1443 &  31 \\
n1316\_gc03147 & 3:22:38.87 & $-$37:19:43 &  22.37 &  1.10 &  1632 &  74 \\
n1316\_gc03151 & 3:22:39.06 & $-$37:15:13 &  21.00 &  1.37 &  1879 &  28 \\
n1316\_gc03151 & 3:22:39.06 & $-$37:15:13 &  21.00 &  1.37 &  1939 &  17 \\
n1316\_gc03161 & 3:22:39.60 & $-$37:15:57 &  21.42 &  1.53 &  1676 &  20 \\
n1316\_gc03192 & 3:22:41.20 & $-$37:20:51 &  21.41 &  1.32 &  1574 &  35 \\
n1316\_gc03199 & 3:22:41.51 & $-$37:18:53 &  23.15 &  1.10 &  1788 &  41 \\
n1316\_gc03269 & 3:22:46.84 & $-$37:14:11 &  21.70 &  1.47 &  1433 &  66 \\
n1316\_gc03280 & 3:22:47.53 & $-$37:14:25 &  21.35 &  1.14 &  1613 &  40 \\
n1316\_gc03314 & 3:22:49.15 & $-$37:11:57 &  21.48 &  1.09 &  1831 &  43 \\
n1316\_gc03318 & 3:22:49.42 & $-$37:11:38 &  20.33 &  1.44 &  1855 &  25 \\
n1316\_gc03327 & 3:22:49.79 & $-$37:09:06 &  22.53 &  1.57 &  1880 &  77 \\
n1316\_gc03332 & 3:22:50.14 & $-$37:13:15 &  22.39 &  1.20 &  2048 & 105 \\
n1316\_gc03336 & 3:22:50.85 & $-$37:12:25 &  20.95 &  1.27 &  1618 &  27 \\
n1316\_gc03338 & 3:22:51.08 & $-$37:12:31 &  22.24 &  1.03 &  1778 &  54 \\
n1316\_gc03351 & 3:22:51.79 & $-$37:11:05 &  21.31 &  1.14 &  1685 &  38 \\
n1316\_gc03374 & 3:22:53.24 & $-$37:10:32 &  21.29 &  1.06 &  1635 &  41 \\
n1316\_gc03381 & 3:22:53.75 & $-$37:11:14 &  21.76 &  1.13 &  1714 &  33 \\
n1316\_gc03394 & 3:22:54.37 & $-$37:13:40 &  23.28 &  1.21 &  2089 &  59 \\
n1316\_gc03397 & 3:22:54.58 & $-$37:11:20 &  21.86 &  1.65 &  1813 &  34 \\
n1316\_gc03411 & 3:22:55.47 & $-$37:09:48 &  22.30 &  1.07 &  1966 &  61 \\
n1316\_gc03417 & 3:22:56.15 & $-$37:11:23 &  21.36 &  1.37 &  1896 &  36 \\
n1316\_gc03422 & 3:22:56.36 & $-$37:09:22 &  21.48 &  1.04 &  1948 &  87 \\
n1316\_gc03451 & 3:22:58.69 & $-$37:14:19 &  21.98 &  1.10 &  1872 &  78 \\
n1316\_gc03541 & 3:23:06.46 & $-$37:14:44 &  22.88 &  1.04 &  1525 &  48 \\
n1316\_gc03580 & 3:23:08.68 & $-$37:06:43 &  23.20 &  0.95 &  1138 &  66 \\
n1316\_gc03632 & 3:23:12.97 & $-$37:09:17 &  21.88 &  1.41 &  1911 &  42 \\
n1316\_gc03667 & 3:23:16.59 & $-$37:15:21 &  21.74 &  1.03 &  1606 &  50 \\
n1316\_gc03723 & 3:23:21.07 & $-$37:10:07 &  22.46 &  0.93 &  1872 &  40 \\
n1316\_gc04101 & 3:22:09.78 & $-$37:15:17 &  22.52 &  1.40 &  1353 &  64 \\
n1316\_gc04128 & 3:22:24.66 & $-$37:15:41 &  22.45 &  1.57 &  1839 &  47 \\
n1316\_gc04128 & 3:22:24.66 & $-$37:15:41 &  22.45 &  1.57 &  1937 &  30 \\
n1316\_gc04132 & 3:22:25.60 & $-$37:10:42 &  22.23 &  1.34 &  1744 &  42 \\
n1316\_gc04138 & 3:22:26.15 & $-$37:17:28 &  22.01 &  1.17 &  1621 &  34 \\
n1316\_gc04138 & 3:22:26.15 & $-$37:17:28 &  22.01 &  1.17 &  1635 &  46 \\
n1316\_gc04146 & 3:22:28.32 & $-$37:15:25 &  23.19 &  1.38 &  1466 &  57 \\
n1316\_gc04149 & 3:22:29.08 & $-$37:14:12 &  21.54 &  1.55 &  1629 &  32 \\
n1316\_gc04150 & 3:22:30.67 & $-$37:14:17 &  20.58 &  1.38 &  1389 &  32 \\
n1316\_gc04155 & 3:22:32.60 & $-$37:16:02 &  21.99 &  1.42 &  1688 &  52 \\
n1316\_gc04224 & 3:22:49.14 & $-$37:09:19 &  22.03 &  1.66 &  2059 &  29 \\
n1316\_gc04238 & 3:22:53.28 & $-$37:10:58 &  21.92 &  1.25 &  2008 &  49 \\
n1316\_gc04518 & 3:22:37.86 & $-$37:18:03 &  20.94 &  1.26 &  1705 &  17 \\
n1316\_gc03579 & 3:23:08.63 & $-$37:04:50 &  21.16 &  1.65 &  1980 &  30 \\
n1316\_gc03505 & 3:23:03.07 & $-$37:08:26 &  19.74 &  1.18 &  1812 &  18\\
n1316\_gc04308 & 3:23:23.05 & $-$37:10:27 &  21.00 &  1.15 &  1887 &  34 \\
n1316\_gc01752 & 3:23:16.64 & $-$37:10:49 &  20.57 &  1.12 &  1771 &  38 \\
n1316\_gc03505 & 3:23:03.13 & $-$37:08:27 &  19.74 &  1.18 &  1749 &  15 \\
n1316\_gc03384 & 3:22:53.89 & $-$37:10:14 &  19.14 &  1.39 &  1976 &  10 \\
n1316\_gc01293 & 3:22:28.88 & $-$37:18:15 &  20.89 &  1.75 &  1825 &  12 \\
n1316\_gc03088 & 3:22:35.23 & $-$37:19:18 &  21.16 &  1.55 &  1838 &  13 \\
n1316\_gc03047 & 3:22:32.64 & $-$37:14:56 &  20.75 &  1.39 &  1607 &  19 \\
\hline
\hline
\label{tab:gcsample_points}
\end{longtable}
%\end{landscape}

%\end{table}

%\resizebox{0.45\textwidth}{!}{
%%\begin{landscape}
\begin{longtable}{ccccccc}
\caption{Identification and radial velocities of globular clusters without entry in the photometric catalog of Paper I.}\\
%\centering
\hline
\hline
Id  &  RA(J2000)  & Dec(J2000) & R & C-R &  rad.vel. &  error \\
\hline
n1316\_gc02855 & 3:21:58.70 & $-$37:12:50 &  23.40 &  1.11 &  1925 &  61 \\
n1316\_gc02965 & 3:22:03.39 & $-$37:14:34 &  22.43 &  1.59 &  1554 &  33 \\
n1316\_gc03030 & 3:22:06.46 & $-$37:08:01 &  22.59 &  1.64 &  1662 &  55 \\
n1316\_gc20839 & 3:22:07.68 & $-$37:05:13 &  22.52 &  1.38 &   821 &  36 \\
n1316\_gc00662 & 3:22:08.47 & $-$37:13:56 &  22.50 &  1.66 &  1708 &  96 \\
n1316\_gc03281 & 3:22:17.03 & $-$37:06:04 &  22.92 &  1.93 &   746 &  50 \\
n1316\_gc20852 & 3:22:19.35 & $-$37:08:53 &  22.72 & -0.03 &  2139 &  64 \\
n1316\_gc20726 & 3:22:26.75 & $-$37:08:42 &  21.96 &  1.77 &  2224 &  64 \\
n1316\_gc00834 & 3:22:27.27 & $-$37:12:54 &  22.40 &  1.41 &  1625 &  24 \\
n1316\_gc00848 & 3:22:29.06 & $-$37:13:48 &  22.71 &  1.42 &  1583 &  51 \\
n1316\_gc10314 & 3:22:29.26 & $-$37:16:11 &  22.95 &  1.24 &  1145 &  55 \\
n1316\_gc11298 & 3:22:34.35 & $-$37:20:29 &  22.84 &  1.50 &  1460 &  76 \\
n1316\_gc20470 & 3:22:34.62 & $-$37:17:01 &  22.57 &  0.90 &  2140 &  63 \\
n1316\_gc20439 & 3:22:37.52 & $-$37:18:58 &  20.83 &  1.52 &  1711 &  20 \\
n1316\_gc20440 & 3:22:38.04 & $-$37:18:46 &  21.71 &  0.22 &  1662 &  20 \\
n1316\_gc20552 & 3:22:41.54 & $-$37:16:30 &  21.97 &  1.27 &  1715 &  30 \\
n1316\_gc20426 & 3:22:42.42 & $-$37:16:48 &  99.99 & 99.99 &  1484 &  26 \\
n1316\_gc04094 & 3:22:46.04 & $-$37:10:28 &  22.55 &  1.73 &   737 &  53 \\
n1316\_gc04140 & 3:22:47.56 & $-$37:10:39 &  22.28 &  1.50 &  1820 &  53 \\
n1316\_gc20237 & 3:22:48.76 & $-$37:10:53 &  22.06 &  1.09 &  1983 &  72 \\
n1316\_gc20282 & 3:22:48.94 & $-$37:14:49 &  21.84 &  1.74 &  1497 &  78 \\
n1316\_gc08412 & 3:22:50.85 & $-$37:12:25 &  20.95 &  1.28 &  1619 & 27 \\ 
n1316\_gc08569 & 3:22:55.53 & $-$37:12:17 &  23.30 &  1.16 &  1510 &  60 \\
n1316\_gc04534 & 3:23:00.26 & $-$37:08:58 &  23.04 &  1.36 &  2023 &  50 \\
n1316\_gc08693 & 3:23:00.35 & $-$37:06:11 &  23.31 &  1.48 &  1821 &  53 \\
n1316\_gc20118 & 3:23:03.76 & $-$37:04:29 &  21.45 &  1.30 &  1976 &  45 \\
n1316\_gc04633 & 3:23:04.61 & $-$37:04:55 &  22.99 &  1.08 &  1318 &  66 \\
n1316\_gc08822 & 3:23:04.84 & $-$37:03:49 &  22.28 &  1.08 &  2917 &  47 \\
n1316\_gc20231 & 3:23:06.55 & $-$37:10:43 &  19.57 &  1.48 &  1596 &   9 \\
n1316\_gc08860 & 3:23:06.70 & $-$37:14:34 &  23.23 &  1.62 &  1212 &  61 \\
n1316\_gc08863 & 3:23:06.75 & $-$37:10:02 &  22.52 &  1.40 &  1661 &  63 \\
n1316\_gc20330 & 3:23:07.69 & $-$37:16:53 &  99.99 & 99.99 &  1798 &  86 \\
n1316\_gc08918 & 3:23:08.98 & $-$37:18:50 &  22.35 &  1.14 &  1774 &  59 \\
n1316\_gc01232 & 3:23:09.96 & $-$37:18:27 &  22.65 &  1.45 &  1255 &  65 \\
n1316\_gc20313 & 3:23:11.47 & $-$37:14:50 &  22.21 &  1.75 &  1773 &  56 \\
n1316\_gc08973 & 3:23:11.72 & $-$37:17:30 &  23.07 &  1.48 &  1563 &  58 \\
n1316\_gc04772 & 3:23:12.22 & $-$37:05:04 &  23.42 &  1.64 &  2322 &  78 \\
n1316\_gc20340 & 3:23:15.02 & $-$37:17:17 &  22.42 &  1.33 &  1950 &  66 \\
n1316\_gc20343 & 3:23:17.95 & $-$37:17:48 &  22.16 &  1.28 &  2473 &  63 \\
n1316\_gc20151 & 3:23:18.46 & $-$37:07:17 &  99.99 & 99.99 &  1078 &  49 \\
n1316\_gc20161 & 3:23:20.33 & $-$37:08:31 &  20.09 &  1.04 &  1888 &  36 \\
n1316\_gc01347 & 3:23:21.56 & $-$37:06:38 &  24.02 &  0.93 &  1564 &  82 \\
\hline
\hline
\normalsize
%}
\label{tab:gcsample}
\end{longtable}
%\end{landscape}

%\begin{table}
%\centering
%\begin{tabular}{ccccccc}
%\hline\hline
%n1316\_gc00250 & 3:22:13.73 & $-$37:11:56.5 &  23.30 &  1.20 &  2020.0 &  61.0 \\
%n1316\_gc00280 & 3:22:23.21 & $-$37:08:26.0 &  22.05 &  5.07 &  1656.1 &  42.9 \\
%n1316\_gc00285 & 3:22:25.89 & $-$37:12:02.1 &  23.50 &  1.32 &  1869.2 &  56.7 \\
%n1316\_gc00286 & 3:22:26.02 & $-$37:05:17.1 &  23.30 &  1.33 &  1427.0 &  49.0 \\
%\hline\hline
%\normalsize
%\end{tabular}
%\label{gcsample}
%\caption{Identification and radial velocities of stars, galaxies, quasars}
%\end{table}

%\section{Sample spectra}
%As an illustration, we show some of our brighter spectra. 

%\bibliographystyle{aa}
%\bibliography{N1316.bib}
%\bibliography{N1316.bib}
\end{document}